\documentclass[aps,prd,10pt,nofootinbib,twocolumn,longbibliography]{revtex4-1}
\usepackage[english]{babel}
\usepackage[autostyle, english = british]{csquotes}
\usepackage[bookmarks=false]{hyperref}
\usepackage{amsmath,amssymb}
\usepackage{graphicx}
\usepackage{enumitem}
\usepackage{hyperref}
\usepackage{xcolor}
\usepackage[T1]{fontenc}
\setlist[enumerate]{topsep=0pt,parsep=-1mm,leftmargin=5mm,}
 \MakeOuterQuote{"}
\usepackage{soul}

\DeclareMathOperator{\arccot}{arccot}
\usepackage{blindtext}

\DeclareMathAlphabet{\mathdutchcal}{U}{dutchcal}{m}{n}
\SetMathAlphabet{\mathdutchcal}{bold}{U}{dutchcal}{b}{n}
\DeclareMathAlphabet{\mathdutchbcal}{U}{dutchcal}{b}{n}

\begin{document}

\title{\large Causal structure of nonhomogeneous dust collapse in effective loop quantum gravity}

\author{Micha{\l} Bobula}
\email{michal.bobula@uwr.edu.pl}
\author{Tomasz Paw{\l}owski}
\email{tomasz.pawlowski@uwr.edu.pl}
\affiliation{University of Wroc{\l}aw, Faculty of Physics and Astronomy, Institute of Theoretical Physics, pl. M. Borna 9, 50-204 Wroc{\l}aw, Poland}

\begin{abstract} 
\noindent We study the causal structure for spherically symmetric dust collapse within a model of effective loop quantum gravity in midisuperspace framework. We develop a general strategy (working beyond the dynamical model of our consideration) for constructing double null coordinates, allowing the extraction of conformal diagrams within single coordinate charts. With the methods introduced, we confirm that the homogeneous Oppenheimer-Snyder collapse scenario resembles the Reissner-Nordstr\"om-like picture. For the nonhomogenous collapse scenario, we construct the conformal diagrams, subsequently, we study its relevant properties, in particular, dust particles' trajectories, apparent horizons and shell-crossing singularities. We conclude that a significant region of spacetime remains inaccessible to the model's dynamics due to the formation of the shell-crossing singularities. The question of whether a timelike singularity, similar to that in the homogeneous dust ball collapse scenario, arises in the nonhomogeneous case remains unresolved. Furthermore, we find that phenomena such as black hole explosions or gravitational shock waves cannot be witnessed by an external observer who does not cross any horizon. Indeed, the collapse cannot take place within single asymptotic region.
\end{abstract}

\maketitle

\section{Introduction}
The problem of gravitational collapse has remained a thorn in the side of General Relativity (GR). Black hole singularities that arise in the matter collapse scenarios indicate a theory breakdown. Even though one cannot presumably witness singularity formation from a distance (as an external observer never crossing the event horizon), the issue can become more relevant by taking into account the black hole evaporation process, as such a process might become an observable phenomenon soon \cite{Coogan:2020tuf}. Indeed, radiating (singular) black holes due to Hawking mechanism \cite{Hawking:1975vcx} develop theoretical difficulties that are known in the literature as the 'information paradox' \cite{Hawking:1976ra, Mathur:2009hf}. For that reason, a consistent picture of black hole formation (and evaporation) is desired --- something that is not provided by GR.  Essentially, singularities of the gravitational collapse are directly responsible for the information puzzle \cite{Ashtekar:2005cj}. Thus, one might expect, that some other framework being an ultraviolet completion of GR or a quantum counterpart of the theory itself could resolve the problems of classical relativity. 

Throughout this work we will focus our attention on effective models of Loop Quantum Gravity (LQG) --- these approaches aim to canonically quantize symmetry-reduced backgrounds and extract corresponding quantum-corrected effective geometries. In this work, we will explore gravitational collapse scenarios within effective LQG and analyze the consequences with the expectation of obtaining a consistent picture of black hole formation as an outcome.

Recently, much effort was devoted to spherically-symmetric dust collapse models in effective LQG \cite{BobulaLOOPS, Bobula:2023kbo, Lewandowski:2022zce, Han:2023wxg, Giesel:2022rxi, Giesel:2023tsj, Giesel:2023hys, Giesel:2024mps, Kelly:2020lec, Kelly:2020uwj, Husain:2021ojz, Husain:2022gwp, Hergott:2022hjm, Fazzini:2023scu,  Fazzini:2023ova, Cipriani:2024nhx, Wilson-Ewing:2024uad, Wilson-Ewing:2024fxo, Munch:2021oqn, Alonso-Bardaji:2023qgu} (see also \cite{Rovelli:2024sjl} for a recent review). In particular, for models based on the so-called improved dynamics scheme \cite{Ashtekar:2006wn} it seems that there is a consensus that a collapse of a homogeneous dust ball exhibits Reissner-Nordstr\"om-like picture of black hole formation. The dust ball bounces and reemerges in a new asymptotic region. What motivates us for conducting this work is the fact that a causal structure of a more realistic, nonhomogeneous case remains unknown, especially the presence of the timelike singularity known from the homogeneous variant is an open issue there. Moreover, the status of the so-called shock formation \cite{Husain:2021ojz, Giesel:2023hys} is unclear. That is, it is advocated that the discontinuities could occur in effective geometry (spacetime metric) when the collapsing matter enters a high energy regime \cite{Husain:2021ojz, Husain:2022gwp} --- this phenomenon might be closely related to formation of the so-called shell-crossing singularities ---  where on some surfaces within probed spacetime regions energy density becomes infinite while geodesic extendibility is maintained \cite{Fazzini:2023ova}. 

The purpose of this work is to carefully analyze the causal structure of models presented in \cite{Kelly:2020lec, Kelly:2020uwj, Husain:2021ojz, Husain:2022gwp, Fazzini:2023ova}. They are built around Hamiltonian description in the Painlev\'e-Gullstrand (PG) gauge for a spherically symmetric background that is a subject of loop quantization procedure. Should the geometry be quantized one obtains effective smooth quantum-corrected geometry by solving effective Ehrenfest-like equations. 

We will confirm, within the model of consideration, that the effective geometry for the homogeneous case is compatible with other similar models where the causal structure is Reissner-Nordstr\"om-like and there are no shock waves in that homogeneous collapse variant. Importantly, we will analyze the nonhomogeneous variant (for a chosen initial Gaussian matter profile). It will turn out that the presence of shell-crossing singularities invalidates the predictions obtained via considered method for a considerable portion of derived geometry. In other words, we will conclude that some spacetime regions cannot be probed with the employed dynamics. We will also address the possible spacetime analytic extensions in a search for the fate of timelike singularity known for the homogeneous case. Moreover, for a nonhomogeneous initial dust configuration, we will examine in greater detail whether an external observer---who does not cross any horizon---can witness purported black hole explosions or gravitational discontinuities, as suggested by the conformal diagrams presented in \cite{Husain:2021ojz, Husain:2022gwp}.

This work will not only explore new physics coming from effective LQG dust collapse models but also develop tools for causal structure analysis, especially those related to Penrose-Carter (conformal) diagram construction. Every diagram presented throughout this work will be numerically computed and displayed within a single coordinate chart. The techniques that will be presented in Sec. \ref{secStrategy} serve as a precursor for a general scheme for deriving a global compactified double null coordinate system that can be applied to any time-dependent spherically symmetric geometry. 

This paper is organized as follows. In Section \ref{secMODEL} we very briefly review the model of our consideration \cite{Kelly:2020lec, Kelly:2020uwj, Husain:2021ojz, Husain:2022gwp, Fazzini:2023ova}, and we also refine some of its aspects for later considerations. In Section \ref{secStrategy} we develop a general strategy for constructing double null coordinates that will be a crucial ingredient for constructions of conformal diagrams studied in following Sections. In Section \ref{secOS} we apply previously proposed tools for studying homogeneous Oppenheimer-Snyder collapse scenario. Subsequently, the main results of this work are presented in Section \ref{secNH} where we analyze the causal structure of the nonhomogeneous dust collapse.

\section{Effective loop quantum gravity dynamics} \label{secMODEL}
\subsection{Model setup}

In setting up the model of our consideration we follow the steps presented in \cite{Husain:2021ojz, Husain:2022gwp}. We consider spherically symmetric dust collapse, where the evolution of the system is governed by an effective Hamiltionian and Brown-Kucha\^r dust field $T$ is the internal clock. The symmetry-reduced background geometry being a subject of loop quantization is fixed to be in 2 relevant gauges, namely Painlev\'e-Gullstrand and areal ones. We write the corresponding line element in Ashtekar-Barbero variables as follows
\begin{equation} \label{metric}
\mathrm{d}\mathdutchcal{s}^2=-\mathrm{d} T^2+\frac{\left(E^b\right)^2}{x^2}\left(\mathrm{d} x+N^x \mathrm{d} T\right)^2+x^2 \mathrm{d} \Omega^2 \, ,
\end{equation}
where $E^b(T,x)$ is triad component being canonically conjugate to Ashtekar-Barbero connection component $b(T,x)$, with (equal $T$-time) Poisson bracket relation
\begin{equation}
    \left\{b\left(x_1\right), E^b\left(x_2\right)\right\}=G \gamma \delta\left(x_1-x_2\right) \,.
\end{equation}
Consequently, the phase space, under the employed gauge choices, is two-dimensional and spanned by the above pair. The areal gauge imposed in the model ensures that the coordinate $x$ is a (metric) radius of 2-spheres ($\mathrm{d} \Omega^2 = \mathrm{d}\theta^2 + \sin^2 \theta \mathrm{d} \varphi^2   $). The shift $N^x$ in \eqref{metric} is given as follows
\begin{equation} \label{shift}
N^x=-\frac{ x}{\gamma \sqrt{\Delta}} \sin \frac{\sqrt{\Delta} b}{x} \cos \frac{\sqrt{\Delta} b}{x} \, ,
\end{equation}
where $\Delta=4 \sqrt{3} \pi \gamma \ell_{\mathrm{Pl}}^2$ is the area gap and $\gamma =0.2375 \dots$ is Barbero-Immirzi parameter (of which value we take the one determined in \cite{Domagala:2004jt, Meissner:2004ju}). The form of the shift vector is fixed by the requirement that the areal gauge is preserved by the dynamics \cite{Kelly:2020uwj, Giesel:2021rky}. The corresponding effective physical Hamiltonian is \cite{Husain:2021ojz, Husain:2022gwp}
\begin{equation} \label{hamiltonian}
\begin{split} 
H_{\text {phys }}^{\text {eff }} & = \int \mathrm{d} x \mathcal{H}_{\text {phys }}^{\text {eff }} = \\ =-\frac{1}{2 G \gamma} & \int \mathrm{d} x\left[\frac{E^b}{\gamma \Delta x} \partial_x\left(x^3 \sin ^2 \frac{\sqrt{\Delta} b}{x}\right)+\frac{\gamma x}{E^b}+\frac{\gamma E^b}{x}\right] \, .
\end{split}
\end{equation}
The time evolution generated by it is captured by the set of Hamilton's equations of motion
\begin{equation} 
    \dot{E^b} = \{ E^b, H_{\text {phys }}^{\text {eff }} \}=-\frac{x^2}{\gamma \sqrt{\Delta}} \partial_x\left(\frac{E_b}{x}\right) \sin  \frac{\sqrt{\Delta} b}{x} \cos \frac{\sqrt{\Delta} b}{x}   \, ,
\end{equation}
\begin{equation} \label{eqb}
\begin{split} 
\dot{b}= \{b, H_{\text {phys }}\} = & \\ =\frac{\gamma x}{2\left(E^b\right)^2}-\frac{\gamma}{2 x}-&\frac{x}{\gamma \Delta} \sin \frac{\sqrt{\Delta}b}{x} \left[\frac{3}{2} \sin \frac{\sqrt{\Delta}b}{x} +x \partial_x \sin \frac{\sqrt{\Delta}b}{x} \right] \, ,
\end{split}
\end{equation}
where dots indicate partial derivatives with respect to $T$. The dust energy density is related to the Hamiltonian as follows \cite{Husain:2021ojz, Husain:2022gwp}
\begin{equation} \label{density}
\rho=-\frac{\mathcal{H}_{\text {phys }}}{4 \pi x\left|E^b\right|} \, .
\end{equation}
The model setup presented in the above paragraphs is an effective description of genuinely quantum one where the gravitational degrees of freedom are subject of loop quantization. The effective equations of motion \eqref{eqb} are in fact Ehrenfest-like equations for the evolution of expectation values of the respective operators living in appropriate Hilbert space -- see \cite{Husain:2021ojz, Husain:2022gwp} for more details.

\subsection{Marginally-trapped dust collapse}

Throughout this work we focus on marginally trapped solutions, that is, metrics describing free-falling dust particles with zero total energy --- their kinetic energy is compensated by the gravitational potential. These metrics correspond to setting $E^b=x$ \cite{Husain:2022gwp}. With that selection the only phase space variable left is $b$ and \eqref{eqb} reduces to
\begin{equation} \label{bdotalpha}
\dot{b}=-\frac{1}{2 \gamma \Delta x} \partial_x\left(x^3 \sin ^2 \frac{\sqrt{\Delta} b}{x}\right) \, .
\end{equation}
We aim to solve the above equation \eqref{bdotalpha}. For that purpose, it is convenient to work with a new function $\beta(T,x):= \sqrt{\Delta} b(T,x)/x$, so that \eqref{bdotalpha} becomes
\begin{equation} \label{betadot}
\dot{\beta}=-\frac{1}{2 \gamma \sqrt{\Delta} x^2} \partial_x\left(x^3 \sin ^2 \beta\right) \, .
\end{equation}
Note that due to the setting $E^b=x$, the line element \eqref{metric} reduces to
\begin{equation} \label{metric2}
\mathrm{d}\mathdutchcal{s}^2 =-\mathrm{d} T^2+\left(\mathrm{d} x+N^x \mathrm{d} T\right)^2+x^2 \mathrm{d} \Omega^2 \, .
\end{equation}
As an aside, notice that for the above line element, $T=const.$ surfaces are flat. Moreover, equations \eqref{hamiltonian}, \eqref{density} combined together allow us to express the energy density with the $\beta$ function as follows
\begin{equation} \label{density2}
\rho=\frac{1}{8 \pi G \gamma^2 \Delta x^2} \partial_x\left(x^3 \sin ^2 \beta \right) \, .
\end{equation}
In particular, on the initial slice $T=0$ slice we can invert \eqref{density2} by integrating the both sides to get
\begin{equation} \label{initial}
\beta^\mathrm{in}(x) =- \arcsin \left(\sqrt{\frac{8 \pi G \gamma^2 \Delta}{x^3} \int_0^x \mathrm{d} \tilde{x} \, \tilde{x}^2 \rho\left(T=0,\tilde{x}\right)}\right) \, .
\end{equation}
The above equation is particularly useful, because later we will see that given some initial density profile $\rho(T=0,x)$, the \eqref{initial} will provide us the corresponding initial condition for $\beta$ at the initial slice. The negative sign was chosen while taking the square root to guarantee that the matter is contracting rather than expanding when evolved (slightly) forward in time from the initial slice.

\subsection{Characteristic equations} \label{subSeccharacteristicslamx0}

We will solve the dynamical equation \eqref{betadot} with the method of characteristics (similarly as was done in original works presenting the model of our consideration \cite{Husain:2021ojz, Husain:2022gwp}). Let $(\lambda, x_0)$ be a pair of parameters on the 2-dimensional surface of solutions to the equation governing model's dynamics \eqref{betadot}. We aim to obtain functions $T(\lambda,x_0)$, $x(\lambda,x_0)$ and $\beta(T(\lambda,x_0), x(\lambda,x_0))$. To achieve that, we choose parameters $\lambda$ and $x_0$ so that they satisfy
\begin{equation} \label{choice}
x(\lambda=0, x_0)=x_0\,, \quad T(\lambda=0, x_0) = 0\,.
\end{equation}
This choice means that $\lambda$ corresponds to $T$ and $x_0$ to $x$ at the initial slice. To write down the characteristic equations, consider a derivative
\begin{equation} \label{fullder}
    \frac{\mathrm{d} \beta}{\mathrm{d}\lambda} =\frac{\partial \beta}{\partial T}  \frac{\mathrm{d} T}{\mathrm{d} \lambda} + \frac{\partial \beta}{\partial x} \frac{\mathrm{d} x}{\mathrm{d}\lambda} \, .
\end{equation}
Now, following the standard techniques for solving partial differential equations we compare coefficients in \eqref{betadot} and \eqref{fullder} to obtain a system of ordinary differential equations for a given $x_0$
\begin{equation} \label{system}
\begin{split}
\frac{\mathrm{d} T}{\mathrm{d} \lambda} = \gamma \sqrt{\Delta}, & \quad \frac{\mathrm{d}x}{\mathrm{d}\lambda} = x(\lambda) \sin \beta(\lambda) \cos \beta(\lambda), \\ \frac{\mathrm{d}\beta}{\mathrm{d}\lambda } &=-\frac{3}{2} \sin^2 \beta(\lambda).
\end{split}
\end{equation}
The initial conditions at the initial slice $T=\lambda=0$ are listed in \eqref{choice} and additionally for $\beta$ we have $\beta(T(\lambda=0, x_0), x(\lambda=0, x_0)) = \beta^{\mathrm{in}}(x_0) $, where $\beta^{\mathrm{in}}$ is given by \eqref{initial}. Next, we find the following solutions
\begin{equation} \label{sol1}
\begin{split}
        T(\lambda, x_0) &= \gamma \sqrt{\Delta} \lambda \,, \\
        x(\lambda, x_0) &=\frac{x_0 \left[ 12 \lambda  \cot \beta^{\mathrm{in}}(x_0)+4 \cot^2\beta^{\mathrm{in}}(x_0)+9 \lambda ^2+4\right]^{ \frac{1}{3}  }}{ ( 4\cot ^2\beta^{\mathrm{in}}(x_0)+4)^{\frac{1}{3}}}\, , \\
        \beta(\lambda, x_0) &= \arccot\left[ \frac{1}{2} \left(3\lambda + 2 \cot \beta^\mathrm{in}(x_0) \right)\right] \,,
\end{split}
\end{equation}
where $\cot$ stands for cotangent function and $\arccot$ is its inverse. Note that with the above solutions, the shift \eqref{shift} can be now expressed as 
\begin{equation} \label{shiftlam}
    N^x (\lambda, x_0) = -\frac{ x(\lambda,x_0) }{\gamma \sqrt{\Delta}} \sin \beta(\lambda, x_0) \cos \beta(\lambda, x_0) \, .
\end{equation}
However, to obtain genuine solution to PDE \eqref{betadot} one has to inverse formulae in \eqref{sol1} to finally write $\beta(T, x)$. That is, one has to look for $\beta$ merely as a function of the original coordinates $(T,x)$. We will demonstrate in next Sections that this procedure (expressing our geometric quantities with $(T,x)$ coordinates) works well for homogeneous dust collapse, but for the nonhomogeneous case it is rather difficult. For that reason, typically we will express our geometric objects with 'new coordinates' $(\lambda, x_0)$ (that we introduced meanwhile discussing method of characteristics) instead of working with\footnote{Keep in mind that the angular coordinates $(\theta, \varphi)$ remain the same in the transformation.} $(T, x)$. Indeed, note that we already have closed, analytic form of our solutions to the characteristic equations \eqref{sol1} --- this could be not the case if we wanted to invert the formulae, especially when we wanted to choose $\beta^\mathrm{in}$ describing nonhomogeneous initial dust profile. 

Besides, note that the solutions \eqref{sol1} yield coordinate transformation $(T, x) \rightarrow (\lambda, x_0)$ for the line element \eqref{metric2}. The transformed metric can be written as
\begin{equation}
\begin{split}
   \mathrm{d}\mathdutchcal{s}^2 = -\gamma ^2 \Delta  \mathrm{d} \lambda^2+\left(\gamma  \sqrt{\Delta } N^x \mathrm{d} \lambda + \frac{\partial x}{\partial \lambda} \mathrm{d} \lambda+ \frac{\partial x}{\partial x_0} \mathrm{d} x_0\right)^2 \\
   + x^2(\lambda,x_0) \mathrm{d} \Omega^2 \, .
\end{split}
\end{equation}
With the use of \eqref{system} and \eqref{shiftlam}, the above line element considerably simplifies to 
\begin{equation} \label{metricLAMX0}
    \mathrm{d}\mathdutchcal{s}^2 = -\gamma ^2 \Delta  \mathrm{d} \lambda^2 + \left(\frac{\partial x}{\partial x_0}\right)^2 \mathrm{d} x_0^2 + x^2(\lambda,x_0) \mathrm{d} \Omega^2 \,.
\end{equation}
The above line element expressed with $(\lambda,x_0)$ coordinates will be particularly useful in the analysis that will be performed in the next sections. We will see that \eqref{metricLAMX0} will exhibit computational advantage over PG coordinates $(T,x)$ and the corresponding line element \eqref{metric2}.

\section{General strategy for double null coordinates} 
\label{secStrategy}

Before we analyse the causal structure of the model under consideration, we will develop a general strategy for constructing double null coordinates. This strategy will let us construct conformal diagrams discussed in the next sections. The methods we are about to present are quite general, that is, the strategy we provide can be utilized for a large class of spherically symmetric metrics. The application of the strategy to the dust collapse in effective LQG will be presented in Sections \ref{secOS} and \ref{secNH} as well as in Fig. \ref{diag_OS}, \ref{diag_OS_ones} and \ref{diag_nonhomo}, \ref{diag_nonhomo_ones}.

To begin with, suppose we work with spherically symmetric line element of the form of
\begin{equation}
\begin{split} \label{general}
    \mathrm{d}\mathdutchcal{s}^2 = & - \mathdutchcal{a}(\mathcal{T, X}) \mathdutchcal{c}(\mathcal{T, X}) \mathrm{d} \mathcal{T}^2 + \mathdutchcal{b}(\mathcal{T}, \mathcal{X}) \mathdutchcal{c}(\mathcal{T, X}) \mathrm{d} \mathcal{T}\mathrm{d} \mathcal{X} \\
   & + \mathdutchcal{a}(\mathcal{T}, \mathcal{X})  \mathdutchcal{e}(\mathcal{T}, \mathcal{X}) \mathrm{d} \mathcal{X} \mathrm{d} \mathcal{T} + \mathdutchcal{b}(\mathcal{T}, \mathcal{X})  \mathdutchcal{e}(\mathcal{T}, \mathcal{X}) \mathrm{d}\mathcal{X}^2 \\ &+ \mathdutchcal{g}(\mathcal{T}, \mathcal{X}) \mathrm{d} \Omega^2 \,,
\end{split}
\end{equation}
where $(\mathcal{T},\mathcal{X})$ are arbitrary coordinates and $\mathdutchcal{a}$, $\mathdutchcal{b}$, $\mathdutchcal{c}$, $\mathdutchcal{e}$, $\mathdutchcal{g}$ are metric components that are known beforehand (for example, they are determined by dynamical equations of the theory of consideration). The metric components in \eqref{general} must satisfy $\mathdutchcal{b} \mathdutchcal{c} = a e$, since the metric must be a symmetric tensor in general. Although one could exclude one of the components from this relation, we retain the form of \eqref{general} for convenience. The line element \eqref{general} might look very specific, however its form reflect the fact that we consider a class of spacetimes whose line elements can be factorized to
\begin{equation} \label{mfactorized}
\begin{split}
   \mathrm{d}\mathdutchcal{s}^2 = \left[ \mathdutchcal{a}(\mathcal{T, X}) \mathrm{d} \mathcal{T} + \mathdutchcal{b}(\mathcal{T}, \mathcal{X}) \mathrm{d}\mathcal{X} \right]& \\
    \times \left[ \mathdutchcal{c}(\mathcal{T, X}) \mathrm{d} \mathcal{T} + \mathdutchcal{e}(\mathcal{T}, \mathcal{X}) \mathrm{d}\mathcal{X} \right]& + \mathdutchcal{g}(\mathcal{T}, \mathcal{X})  \mathrm{d} \Omega^2
\end{split}
\end{equation}
Indeed, the method will simply work for metrics viable for the above factorization. The radial null condition ($\mathrm{d}\mathdutchcal{s}^2=0$ and $\mathrm{d}\Omega =0$) is given by vanishing of either the first or the second factor in \eqref{mfactorized}. We write the differentials of the double null coordinates we are about to construct, say $(u, v)$, as the null conditions multiplied by some functions $f^u$ and $f^v$ respectively
\begin{equation} \label{dnStrategy}
\begin{split}
    \mathrm{d}u &= f^u(\mathcal{T}, \mathcal{X}) \left[ \mathdutchcal{a}(\mathcal{T, X}) \mathrm{d} \mathcal{T} + \mathdutchcal{b}(\mathcal{T}, \mathcal{X}) \mathrm{d}\mathcal{X} \right] \,, \\
    \mathrm{d}v &= f^v(\mathcal{T}, \mathcal{X}) \left[  \mathdutchcal{c}(\mathcal{T, X}) \mathrm{d} \mathcal{T} + \mathdutchcal{e}(\mathcal{T}, \mathcal{X}) \mathrm{d}\mathcal{X} \right] \,.
\end{split}
\end{equation}
The functions  $f^u$ and $f^v$, introduced as a necessary ingredient of the construction of the double null coordinates, will be extracted as follows: First we note one key property of the forms above --- by definition the differential forms $\mathrm{d}u$ and $\mathrm{d}v$ must be exact, thereby, their exterior derivative must be zero
\begin{equation} \label{exact}
\mathrm{d}(\mathrm{d}u)=0\, \quad \mathrm{d}(\mathrm{d}v)=0\,.
\end{equation}
This requirement arises from general properties of differential forms on manifolds --- see for example the discussion around Eqs.~14.17--14.18 in \cite{Lee2013}.

In terms of the component functions the conditions \eqref{exact} take the form
\begin{equation}
\begin{split}
    \frac{\partial \left( f^u\,\mathdutchcal{a} \right) }{\partial \mathcal{X}} \mathrm{d} \mathcal{X} \wedge \mathrm{d} \mathcal{T} +  \frac{\partial \left( f^u\,\mathdutchcal{b} \right) }{\partial \mathcal{T}} \mathrm{d} \mathcal{T} \wedge \mathrm{d} \mathcal{X} = 0\,, \\
    \frac{\partial \left( f^v\,\mathdutchcal{c} \right) }{\partial \mathcal{X}} \mathrm{d} \mathcal{X} \wedge \mathrm{d} \mathcal{T} +  \frac{\partial \left( f^v\,\mathdutchcal{e} \right) }{\partial \mathcal{T}} \mathrm{d} \mathcal{T} \wedge \mathrm{d} \mathcal{X} = 0\,.
\end{split}
\end{equation}
Which holds iff $f^u$ and $f^v$ satisfy the following partial differential conditions 
\begin{equation} \label{dbPDE}
\begin{split}
    \frac{\partial \left( f^u\,\mathdutchcal{a} \right) }{\partial \mathcal{X}} - \frac{\partial \left( f^u\,\mathdutchcal{b} \right) }{\partial \mathcal{T}} = 0\,, \\
      \frac{\partial \left( f^v\,\mathdutchcal{c} \right) }{\partial \mathcal{X}} - \frac{\partial \left( f^v\,\mathdutchcal{e} \right) }{\partial \mathcal{T}} = 0\,.
\end{split}
\end{equation}
The equations \eqref{dbPDE} require additional discussion. They are a relatively simple 1st order system, thus it is easy to formulate appropriate initial/boundary value problem guaranteeing existence of their unique solution, however the values of $f^u,f^v$ at initial/boundary surface are free. This translates into a huge freedom of constructing the coordinates $u,v$ and there is no simple choice restricting this freedom. It is however possible to make choices particularly convenient for certain applications, which we will demonstrate further.

The above freedom has a consequence for the appearance of the Penrose-Carter diagram, as the one constructed with use of one solution to \eqref{dbPDE} appears distorted with respect to the diagram constructed with use of another solution. A good example of that are the diagrams in Figures \ref{diag_OS} and \ref{diag_OS_ones}. Note however, that the freedom of a null coordinate choice does not affect the causal structure encoded in the diagram\footnote{When two Penrose-Carter diagrams have exactly the same causal structure but look different, it means they represent the same underlying relationships between events in spacetime—specifically, which events are causally separated or connected—despite being depicted differently. If two diagrams preserve all of these relationships then they are said to have the same causal structure. The reason they might look different is because Penrose-Carter diagrams are conformal diagrams and the coordinates in which they are drawn are not unique. The only elements that are invariant are the null lines always being at 45-degree angle, and the relations as to whether one point of spacetime is above/below/outside of a (n invariant on the diagram) null cone of another. The exact shapes of the interiors of ``diamonds'' distinguished by null geodesics as well as exact parametrizations of these null geodesics can be deformed by diffeomorphisms.}.

Having $f^u$ and $f^v$ determined from \eqref{dbPDE} we are now ready to write the double null coordinates as
\begin{equation} \label{uvgen}
\begin{split}
  u=  \int f^u(\mathcal{T}, \mathcal{X}) \mathdutchcal{a}(\mathcal{T, X}) \mathrm{d} \mathcal{T} + \int f^u(\mathcal{T}, \mathcal{X}) \mathdutchcal{b}(\mathcal{T}, \mathcal{X}) \mathrm{d}\mathcal{X}\,, \\
    v=  \int f^v(\mathcal{T}, \mathcal{X}) \mathdutchcal{b}(\mathcal{T, X}) \mathrm{d} \mathcal{T} + \int f^v(\mathcal{T}, \mathcal{X}) \mathdutchcal{e}(\mathcal{T}, \mathcal{X}) \mathrm{d}\mathcal{X}.
\end{split}
\end{equation}
However, the coordinates $(u, v)$ constructed above might not necessarily be compact or cover the whole region of spacetime we are interested in. Therefore, one has to specifically compactify them in order to construct a conformal diagram covering the whole region of interest. For that we propose the following transformation
\begin{equation} \label{comp}
    \begin{split}
        \tilde{u}(u)= \pm 1/\pi\arctan u + c_u \, , \\
        \tilde{v}(v)= \pm 1/\pi \arctan v + c_v \, .
    \end{split}
\end{equation}
where $c_u, c_v \in \{\dots, -1, 0, 1, \dots \}$ are constants indicating position of particular diagram's \emph{block}\footnote{A precise definition of the block can be found in \cite{Schindler:2018wbx}. However, for the sake of an example one can think about maximally extended Schwarzschild spacetime where there are four blocks separated by the horizons --- black hole exterior, black hole interior, parallel black hole exterior and white hole interior. }. One has to switch between plus and minus sign in \eqref{comp} when crossing each diagram's block --- these crossing points corresponds to situations when coordinates $u$ or $v$ diverge and sign change preserves monotonicity of $\tilde{u}$ and $\tilde{v}$. We will elaborate on this issue in the context of particular applications of the above method in Sections \ref{secOS} and \ref{secNH}.

\section{Oppenheimer-Snyder collapse scenario}
\label{secOS}

With the dynamical equations and tools for double null coordinates construction established in Sections \ref{secMODEL} and \ref{secStrategy}, we now can analyze the causal structure for homogeneous dust ball collapse. For that, we will study modified Oppenheimer-Snyder collapse scenario within effective LQG. As stated before, we aim to analyze the causal structure of the model of the consideration that was originally introduced in \cite{Husain:2021ojz, Husain:2022gwp}. To achieve our objective, we will work both with coordinates $(\lambda, x_0)$ and $(T,x)$ discussed in Subsection \ref{subSeccharacteristicslamx0}. More precisely, we will express our yet-to-be-constructed double null coordinates as functions $u(\lambda, x_0)$, $v(\lambda, x_0)$, however some discussions regarding the spacetime properties will be performed with use of $(T,x)$ coordinates. The reason to use $(\lambda, x_0)$ coordinates is the fact that some calculations, especially a construction of abovementioned double null coordinates, will be considerably simplified with them.

\subsection{Metric components in Painlev\'e-Gullstrand coordinates}

For the collapse of the homogeneous dust ball (known as the Oppenheimer-Snyder collapse scenario) we write the dust energy density profile at the initial constant time slice as
\begin{equation} \label{rhoin}
    \rho_{\mathrm{OS}}(T=0,x) = \begin{cases} \rho_0 \;\; \mathrm{for} \;\; x \leq x_b\,,\\ 0 \;\; \mathrm{for} \;\; x>x_b\,.\end{cases} 
\end{equation}
where $x_b$ is a chosen radius of the dust ball at the initial slice $T=\lambda=0$. Note that the interior dust ball region is given by $x\leq x_b$ and, conversely, the exterior vacuum is given by $x>x_b$ at the initial slice. Based on the method of the characteristic equations discussed in Section \ref{secMODEL}, we aim to determine the geometry (metric components) for the studied homogenous dust collapse. With the choice of the initial energy density \eqref{rhoin}, $\beta^\mathrm{in}$ in \eqref{initial} reduces to
\begin{equation} \label{betainOS}
    \beta^\mathrm{in}_\mathrm{OS}(x_0) = \begin{cases} -\arcsin \left(2 G M \gamma^2 \Delta/ x_b^3\right)^{1/2} \;\; \mathrm{for} \;\; x \leq x_b \,, \\
    -\arcsin \left(2 G M \gamma^2 \Delta/ x_0^3\right)^{1/2} \;\; \mathrm{for} \;\; x>x_b\,,
    \end{cases}
\end{equation}
where we define $M:= 4/3 \pi x_b^3 \rho_0 $. One has to plug the expression \eqref{betainOS} to general solutions of the characteristic equations \eqref{sol1} in order to derive specific solutions for the homogeneous scenario with the initial profile \eqref{rhoin} we have chosen. Indeed, to complete the  method of characteristics we aim to express the function $\beta(\lambda, x_0)$ in \eqref{sol1} as merely function of the original PG coordinates $\beta(T, x)$. To perform this, one has to find inverses of our characteristics \eqref{sol1} to write $\beta(\lambda(T, x), x_0(T, x) ) = \beta(T,x)$. The result of the computation is 
\begin{equation} \label{betasolOSxT}
    \beta_\mathrm{OS}(T,x) = \begin{cases}
        \arctan \frac{2 \gamma \sqrt{\Delta}}{3 T - \gamma \sqrt{\frac{2 x_b^3}{G M \gamma^2}-4\Delta}} \;\;  \text{for the interior}\,, \\
        \arctan \frac{\gamma}{\gamma^2 - \frac{x^3}{2 G M \Delta}} \;\; \text{for the exterior}\,.
    \end{cases}
\end{equation}
The procedure of solving \eqref{betadot} with the method of characteristics is now completed. The geometry for the Oppenheimer-Snyder collapse scenario is uniquely determined thanks to \eqref{betasolOSxT}. Indeed, note that with the above formula \eqref{betasolOSxT}, we can express the shift \eqref{shiftlam} as
\begin{equation} \label{shiftOS}
        N^x_\mathrm{OS}(T,x)= \begin{cases}
        \begin{split}
            \frac{2x\left( -3 G M T + \sqrt{2 G M \left(x_b^3 - 2 G M \gamma^2 \Delta\right)}\right)}{9 GM T^2 + 2 x_b^3 - 6 T \sqrt{2G M \left(x_b^3 - 2 G M \gamma^2 \Delta\right) }} \\
            \;\; \text{for the interior}\,,
        \end{split}\\
        \sqrt{\frac{2 G M \left(x^3 - 2 G M \gamma^2 \Delta\right)}{x^4}}  \;\; \text{for the exterior}\,.
    
        \end{cases}
\end{equation}
The derived geometry for the interior of the dust ball can be put into more familiar form by introducing a new coordinate $r$ for the line element \eqref{metric2} defined via a relation
\begin{equation} \label{scalex}
a(T)r = x\;\; \text{for the interior} \,,
\end{equation}
where function $a$ will be defined in a moment (in fact, $a$ will be a cosmological scale factor). First, note that equivalently we can write \eqref{scalex} as 
\begin{equation} \label{adr}
    a(T) \mathrm{d}r = \mathrm{d}x - \frac{\dot{a}}{a} x \, \mathrm{d}T  \,.
\end{equation}
Second, for the line element \eqref{metric2} in the interior of the dust ball we can identify
\begin{equation} \label{Nxdotaa}
N^x_\mathrm{OS}=-\frac{\dot{a}}{a} x \,.
\end{equation}
This, in turn, with the use of \eqref{shiftOS}, the \eqref{Nxdotaa}  can be integrated to obtain
\begin{equation} \label{scaleEx}
    a(T) = \left( 9 G M T^2 + 2 x_b^3 - 6 T \sqrt{2 G M \left(x_b^3 -2GM\gamma^2 \Delta\right)} \right)^{1/3} \,.
\end{equation}
Finally, we can combine together \eqref{adr} and \eqref{Nxdotaa} to transform the line element  \eqref{metric2} into
\begin{equation} \label{frw}
    \mathrm{d}\mathdutchcal{s}^2 = - \mathrm{d} T^2 + a^2(T) \mathrm{d}r^2 + a^2(T)r^2 \mathrm{d}\Omega^2 \,,
\end{equation}
in the interior. With the scale factor of the form \eqref{frw} the metric \eqref{scaleEx} is in facto that of a bouncing FRW geometry known from loop quantum cosmology. 

Subsequently, for the exterior of the dust ball we employ Schwarzchild-like time $t$ defined as follows
\begin{equation}
    \mathrm{d}t = \mathrm{d}T- \frac{1}{1-(N^x)^2} N^x \mathrm{d} x \, ,
\end{equation}
so that the line element \eqref{metric2} can be recasted to
\begin{equation}
    \mathrm{d}\mathdutchcal{s}^2 = -(1-(N^x)^2)\mathrm{d}t + \frac{1}{1-(N^x)^2} \mathrm{d}x + x^2 \mathrm{d}\Omega^2 \,,
\end{equation}
where
\begin{equation} \label{schwarz}
    1-(N^x_\mathrm{OS})^2 = 1 - \frac{2 G M}{x} + \frac{4 G^2 M^2 \gamma^2 \Delta}{x^4} \;\; \text{for the exterior} \,.
\end{equation}
We can see that \eqref{schwarz} yield a modified Schwarzschild geometry --- the first two terms represent the classical part, however, the third one plays the role of the quantum correction. This outcome was also obtained within the model of the consideration in \cite{Husain:2021ojz, Husain:2022gwp}. With different methods based on gluing cosmological metrics (determined by Loop Quantum Cosmology) with spherically symmetric vacua where the exterior is completely determined by Israel-Darmois junction conditions, \eqref{scaleEx} and \eqref{schwarz} were derived in \cite{BobulaLOOPS,Bobula:2023kbo, Lewandowski:2022zce}.

\subsection{Dust particles' trajectories and apparent horizons} 
\label{subsecParttraj}

Before we construct the double null coordinates, let us study relevant geometric objects that we will later display on the conformal diagram. So far, we know that the initial slice is given by $\lambda = 0$ (equivalently $T=0$), and $x_0$ is a parameter on that slice. We want to determine radial trajectories of free-falling dust particles (or timelike observers).  In PG coordinates, consider an integral curve $y^\alpha(T)=(T, x=R(T))$ with respect to the line element \eqref{metric2} where the shift can have a general form $N^x(T,x)$. The tangent is $l^\alpha:=\frac{\mathrm{d} y^\alpha}{\mathrm{d}T} = (1, R'(T))$, and from the normalization condition $l^\alpha l_\alpha = -1$ we immediately have $R'(T) = -N^x$. Thus, the tangent satisfies
\begin{equation}
    l^\alpha \nabla_\alpha l^\beta =0 \,.
\end{equation}
Indeed, the four-acceleration is zero --- it means that $T$ is proper time on radial timelike geodesics, as expected for PG coordinates. Since from the solutions \eqref{sol1} we learn that $\lambda$ is proportional to $T$, $\lambda$ is also an affine parameter on these geodesics. That is, the corresponding integral curves can be parametrized as $y^\alpha(T(\lambda))$ and $x_0$ labels each geodesic from the congruence originating from the initial surface. Note that since $x_0$ remains constant the characteristic curves being solutions to equations \eqref{system}, this parameter will also remain constant on each timelike geodesic discussed above. To understand this issue more clearly, one can, alternatively, start with the line element \eqref{metricLAMX0} written with coordinates $\left(\lambda, x_0 \right)$ (where $N^x(\lambda, x_0)$ has a general form) and then write a timelike integral curve as $z^\alpha (\lambda) = (\lambda, x_0 = const. )$. Straightforward calculation gives
\begin{equation}
    \frac{\mathrm{d}z^\alpha}{\mathrm{d} \lambda } \nabla_\alpha \frac{\mathrm{d}z^\beta}{\mathrm{d} \lambda } =0 \,,
\end{equation}
which confirms the above statements --- the four acceleration is zero for the $\lambda$-parametrization where $x_0 = const.$ on each geodesic. 

One of the critical objects of interests are the apparent horizons that form during the gravitational collapse. To detect them we will formulate conditions for vanishing of radial null geodesics' expansions. Let $T$ be a parameter on null geodesics. The line element \eqref{metric2} delivers two distinct null conditions ($\mathrm{d}\mathdutchcal{s}^2=0$) for radial ingoing and outgoing null geodesics. Hence, tangents to ingoing geodesics can be written as $k^\alpha_{\mathrm{ing}}=\frac{\mathrm{d} x^\alpha}{\mathrm{d}T} = (1, -1-N^x(T,x)) $, and $k^\alpha_{\mathrm{out}}= \frac{\mathrm{d} x^\alpha}{\mathrm{d}T} = (1,1-N^x(T,x)) $ for the outgoing geodesics. The tangent vector field for the first family, that is, ingoing null geodesics, satisfies
\begin{equation}
    \begin{split}
        k^\alpha_{\mathrm{ing}} \nabla_\alpha k^\beta_\mathrm{ing} = \kappa_\mathrm{ing} k^\beta_\mathrm{ing}\,, \quad \kappa_\mathrm{ing} = -\partial_x N^x(T,x)\,,
    \end{split}
\end{equation}
and similarly for the outgoing family we have $\kappa_\mathrm{out} = \kappa_\mathrm{in} = -\partial_x N^x(T,x)$. Apparently, $T$ is not an affine parameter on these null geodesics. Should the geodesics be affinely parametrized, say with the affine parameter $T^*$, the expansions would be given by formulae $\vartheta_\mathrm{ing} = \nabla_\alpha k^\alpha_{\mathrm{ing}*}$ and $\vartheta_\mathrm{out} = \nabla_\alpha k^\alpha_{\mathrm{out}*}$ (see for example Chapters 2.4 and 5.1.8 of \cite{Poisson:2009pwt} for a general discussion). However, to compute these quantities we need to relate the parameter $T$ with the affine one $T^*$. They can be related to each other by the condition (see Chapter 1.3 of \cite{Poisson:2009pwt})
\begin{equation}
    \frac{\mathrm{d}^2T^*}{\mathrm{d}T^2 } = -\partial_x N^x(T,x) \frac{\mathrm{d}T^*}{\mathrm{d}T}\,.
\end{equation}
Solving this differential condition we get $\frac{\mathrm{d}T^*}{\mathrm{d}T} = e^{-\Gamma} $, where $\frac{\mathrm{d \Gamma}}{\mathrm{d}T} = -\partial_x N^x(T,x) $. The abovementioned expansion then takes the form
\begin{equation} \label{expaa}
    \vartheta_\mathrm{ing} = \nabla_\alpha k^\alpha_{\mathrm{ing}*} = \nabla_\alpha \left(e^{-\Gamma} k^\alpha_{\mathrm{ing}} \right) = e^{-\Gamma} \left( \nabla_\alpha k^\alpha_{\mathrm{ing}}  -  \kappa_\mathrm{ing} \right) \,.
\end{equation}
By the same arguments an analogous formula holds also for the expansion of the radial outgoing null geodesics $\vartheta_\mathrm{out}$. From the straightforward calculation, we further have
\begin{equation}
\begin{split}
    \nabla_\alpha k^\alpha_{\mathrm{ing}} &= -\frac{1}{x} \left( 2 + 2 N^x(T,x)+ x \partial_x N^x(T,x) \right) \,, \\
        \nabla_\alpha k^\alpha_{\mathrm{out}} &= \frac{1}{x} \left( 2 -2 N^x(T,x) - x \partial_x N^x(T,x) \right) \,.
\end{split}
\end{equation}
Note that according to \eqref{expaa}, the expansions $\vartheta_\mathrm{ing}$ and $\vartheta_\mathrm{out}$ vanish whenever $\nabla_\alpha k^\alpha_{\mathrm{ing}}  -  \kappa_\mathrm{ing}$ and $\nabla_\alpha k^\alpha_{\mathrm{out}}  -  \kappa_\mathrm{out}$ are zero respectively. Thanks to the computations performed in the above discussion, these conditions can be simplified to finding roots of $1+N^x(T,x)$ for the ingoing geodesics and $1-N^x(T,x)$ for outgoing ones. Equivalently, the apparent horizons are located at zeros of $1+N^x(\lambda,x_0)$ and $1-N^x(\lambda,x_0)$ where the shift $N^x$ could be specifically given by \eqref{shiftlam}.

\subsection{Double null coordinates}

Applying the tools presented in Section \ref{secStrategy} we construct suitable double null coordinates so that we will build the conformal diagram for Oppenheimer-Snyder collapse scenario. The starting point for us is the line element \eqref{metricLAMX0} written with coordinates $(\lambda, x_0)$. There are two null conditions ($ \mathrm{d}\mathdutchcal{s}^2 =0$) with respect to \eqref{metricLAMX0} letting us to identify $\mathcal{T} \rightarrow \lambda$, $\mathcal{X}\rightarrow x_0 $ and to rewrite \eqref{dnStrategy} as
\begin{equation} \label{nulllamx0}
    \begin{split}
    \mathrm{d}u & \rightarrow f^u(\lambda, x_0) \left( -\gamma \sqrt{\Delta} \mathrm{d} \lambda + \frac{\partial x}{ \partial x_0} \mathrm{d}x_0 \right)  \,, \\
    \mathrm{d}v & \rightarrow f^v(\lambda, x_0) \left( -\gamma \sqrt{\Delta} \mathrm{d} \lambda - \frac{\partial x}{ \partial x_0} \mathrm{d}x_0 \right) \,.
\end{split}
\end{equation}
Next, we compute the conditions for closeness of $\mathrm{d}u,\mathrm{d}v$: $\mathrm{d} (\mathrm{d}u) =0 $ and $\mathrm{d} (\mathrm{d} v) =0$. They yield
\begin{equation}
    \begin{split}
        \frac{\partial}{\partial x_0} &\left( -\gamma \sqrt{\Delta} f^u(\lambda,x_0) 
 \right)  \mathrm{d} x_0 \wedge \mathrm{d} \lambda \\ &+ \frac{\partial }{\partial \lambda} \left( f^u(\lambda,x_0)  \frac{\partial x}{ \partial x_0} \right)  \mathrm{d} \lambda \wedge \mathrm{d} x_0 =0 \,,
    \end{split}
\end{equation}
\begin{equation}
    \begin{split}
        \frac{\partial}{\partial x_0} & \left( -\gamma \sqrt{\Delta} f^v(\lambda,x_0) 
 \right) \mathrm{d} x_0 \wedge \mathrm{d} \lambda \\ &+ \frac{\partial }{\partial \lambda} \left( -f^v(\lambda,x_0)  \frac{\partial x}{ \partial x_0} \right) \mathrm{d} \lambda \wedge \mathrm{d} x_0 =0 \,.
    \end{split}
\end{equation}
Thus, the functions $f^u$ and $f^v$ have to satisfy
\begin{equation} \label{rowf1}
        \frac{\partial f^u(\lambda, x_0)}{\partial \lambda} \frac{\partial x}{\partial x_0} + f^u(\lambda, x_0) \frac{\partial^2 x }{\partial \lambda \partial x_0} + \gamma \sqrt{\Delta} \frac{\partial f^u(\lambda, x_0)}{\partial x_0} =0 \,, 
        \end{equation}
        \begin{equation} \label{rowf2}
               \frac{\partial f^v(\lambda, x_0)}{\partial \lambda} \frac{\partial x}{\partial x_0} + f^v(\lambda, x_0) \frac{\partial^2 x }{\partial \lambda \partial x_0} - \gamma \sqrt{\Delta} \frac{\partial f^v(\lambda, x_0)}{\partial x_0} =0 \,,
\end{equation}
respectively. To find the double null coordinates we need to solve \eqref{rowf1} and \eqref{rowf2}. We again employ the method of characteristic equations to solve PDEs in \eqref{rowf1} and \eqref{rowf2}. Let a pair $(\lambda^*, s)$ be parameters on 2-dimensional surfaces of solutions to \eqref{rowf1} and \eqref{rowf2}. We look for expressions $\lambda(\lambda^*,x)$, $x_0(\lambda^*,x)$ and $f^u(\lambda(\lambda^*,s), x_0(\lambda^*,s))$ for the case of to-be-extracted null $u$ coordinate and a similar triple for the case of $v$ coordinate (where we look for $f^v$ instead). For that let us consider the following derivatives
\begin{equation} \label{derfu}
    \frac{\mathrm{d} f^u}{\mathrm{d}s} = \frac{\partial f^u}{\partial \lambda} \frac{\mathrm{d} \lambda}{ \mathrm{d} s} + \frac{\partial f^u}{\partial x_0} \frac{\mathrm{d} x_0}{ \mathrm{d} s} \,,
\end{equation}
and
\begin{equation}  \label{derfv}
       \frac{\mathrm{d} f^v}{\mathrm{d}s} = \frac{\partial f^v}{\partial \lambda} \frac{\mathrm{d} \lambda}{ \mathrm{d} s} + \frac{\partial f^v}{\partial x_0} \frac{\mathrm{d} x_0}{ \mathrm{d} s} \,.
\end{equation}
Again, following the standard procedure of the characteristic equations we compare \eqref{derfu} to \eqref{rowf1} and \eqref{derfv} to \eqref{rowf2}. These comparisons allows us to write characteristic equations for $f^u$ as (we also pick initial conditions)
\begin{equation} \label{systemSstarU}
    \begin{split}
        &\frac{\mathrm{d} \lambda}{\mathrm{d} s} = \frac{\partial x}{\partial x_0} \,, \quad \lambda(s=0) = \lambda^* \,, \\ &\frac{\mathrm{d} x_0}{\mathrm{d} s} = \gamma \sqrt{\Delta} \,, \quad x_0(s=0)=x_{\mathrm{in}, s=0}^u \,, \\
        &\frac{\mathrm{d} f^u }{\mathrm{d} s} = -f^u \frac{\partial^2 x}{\partial x_0 \partial \lambda} \,, \quad f^u(s=0) = f^u_{\mathrm{in}, s=0} \,.
    \end{split}
\end{equation}
Similarly, for $f^v$ we have
\begin{equation} \label{systemSstarV}
    \begin{split}
       & \frac{\mathrm{d} \lambda}{\mathrm{d} s} = \frac{\partial x}{\partial x_0} \,, \quad \lambda(s=0) = \lambda^* \,, \\& \frac{\mathrm{d} x_0}{\mathrm{d} s} =- \gamma \sqrt{\Delta} \,, \quad x_0(s=0)=x_{\mathrm{in}, s=0}^v \,, \\
        &\frac{\mathrm{d} f^v }{\mathrm{d} s} = -f^v \frac{\partial^2 x}{\partial x_0 \partial \lambda} \,, \quad f^v(s=0) = f^v_{\mathrm{in}, s=0} \,,
    \end{split}
\end{equation}
where $x_{\mathrm{in}, s=0}^u$, $f^u_{\mathrm{in}, s=0}$ and $x_{\mathrm{in}, s=0}^v$, $f^v_{\mathrm{in}, s=0}$ are the (free) initial conditions --- they need to be specified (chosen). As stated in Section \ref{secStrategy} we did not provide any algorithm for specifying them. Instead, we choose them based on some reasonable expectations. The choice will determine the general appearance of the conformal diagram --- for example, see the difference between the diagram in Figure \ref{diag_OS} and the one in Figure \ref{diag_OS_ones} where the choices were different on each of them. To extract 'traditionally' looking diagram, e.g. the one in Figure \ref{diag_OS}, we choose
\begin{equation} \label{in1}
\begin{split}
x_{\mathrm{in}, s=0}^u \rightarrow x_b\,,\;\; f^u_{\mathrm{in}, s=0}& \rightarrow -1/[1-N^x(\lambda = \lambda^*, x_0 = x_b) ]\,, \\
x_{\mathrm{in}, s=0}^v \rightarrow x_b \,, \;\; f^v_{\mathrm{in}, s=0} &\rightarrow 1/[1+N^x(\lambda = \lambda^*, x_0 = x_b) ] \,.
\end{split}
\end{equation}
This choice will make the coordinates $u, v$ resemble the typical double null coordinates for black hole spacetimes (like the coordinates (5.5) in \cite{Poisson:2009pwt}). Besides, we also consider a relevantly different choice of initial conditions, namely 
\begin{equation} \label{in2}
\begin{split}
x_{\mathrm{in}, s=0}^u \rightarrow x_b\,,\;\; f^u_{\mathrm{in}, s=0}& \rightarrow -1\,, \\
x_{\mathrm{in}, s=0}^v \rightarrow x_b \,, \;\; f^v_{\mathrm{in}, s=0} &\rightarrow 1 \,.
\end{split}
\end{equation}
This choice will yield the conformal diagram in the Figure \ref{diag_nonhomo_ones}. The appearance of the resulting diagram is rather non-standard, however, the causal structure will be exactly the same as in the case of traditional one in Figure \ref{diag_OS}, as expected. The initial conditions \eqref{in2} are advantageous over \eqref{in1}, because they will give a coordinate system that can be more easily compactified while covering the whole spacetime of our study. In particular, more advantages of the choice similar to \eqref{in2} will become apparent in the case of nonhomogeneous dust collapse studied in Section \ref{secNH}.

To finish the extraction of desired functions $f^v$ and $f^u$ after the above steps, one should first solve the systems of equations \eqref{systemSstarU} and \eqref{systemSstarV}. We solve the systems numerically utilizing libraries of \textsc{JULIA} programming language. The second step, would be to look for inverses $s(\lambda, x_0)$ and $\lambda^*(\lambda, x_0)$ to write e.g $f^u(\lambda, x_0) = f^u(\lambda^*(\lambda, x_0), s(\lambda, x_0))$ (analogously for $f^v$). However, we will slightly modify the general strategy presented in Section \ref{secStrategy}. That is, we will rewrite \eqref{nulllamx0} with $(\lambda^*,s )$ coordinates. We perform this transformation, because it would considerably simplify the integration of $\mathrm{d}u$ and $\mathrm{d}v$ (see \eqref{uvgen} for the general expressions). Hence, \eqref{nulllamx0} rewritten in $(\lambda^*, s)$ coordinates is (note that \eqref{systemSstarU} and \eqref{systemSstarV} determine the transformation rules $(\lambda, x_0) \rightarrow (\lambda^*, s)$)
\begin{equation}
    \begin{split}
        \mathrm{d} u &\rightarrow f^u(\lambda^*, s) \left( - \gamma \sqrt{\Delta} \frac{\partial \lambda}{\partial \lambda^*} + \frac{\partial x_0 }{\partial \lambda^*} \frac{\partial x}{\partial x_0}  \right) \mathrm{d} \lambda^* \,, \\
        \mathrm{d} v &\rightarrow f^v(\lambda^*, s) \left( - \gamma \sqrt{\Delta} \frac{\partial \lambda}{\partial \lambda^*} - \frac{\partial x_0 }{\partial \lambda^*} \frac{\partial x}{\partial x_0}  \right) \mathrm{d} \lambda^* \,.
    \end{split}
\end{equation}
Indeed, $\mathrm{d}u$ and $\mathrm{d}v$ are independent of $\mathrm{d}s$. It means that the values of $u$ and $v$ at say $s=0$ remain constant along the characteristic curves (parametrized by $s$). We choose the following conditions for $u$ and $v$
\begin{equation}
    \begin{split}
        u(\lambda^*=0, s=0) =0 \,, \quad v(\lambda^*=0, s=0) =0 \,.
    \end{split}
\end{equation}
Hence, we can write the following
\begin{equation} \label{nullamstars}
\begin{split} 
    u(\lambda^*, s) = -\gamma \sqrt{\Delta} \int^{\lambda^*}_0 f^u(\lambda^*,s=0) \,\mathrm{d}\lambda^*\,, \\
    v(\lambda^*, s) = -\gamma \sqrt{\Delta} \int^{\lambda^*}_0 f^v(\lambda^*,s=0) \,\mathrm{d}\lambda^*\,,
    \end{split}
\end{equation}
where we have used the relations $\frac{\partial x_0}{\partial \lambda^*} =0$ and $\frac{\partial \lambda }{\partial \lambda^*} =1 $ holding at $s=0$. Finally, by numerically finding the inverses $s(\lambda, x_0)$ and $\lambda^*(\lambda, x_0)$ we can express the above double null coordinates \eqref{nullamstars} as functions of $(\lambda, x_0)$
\begin{equation} \label{uvlamx0_f}
    \begin{split}
        u(\lambda, x_0) &= u(\lambda^*(\lambda, x_0), s(\lambda, x_0)) \,, \\
        v(\lambda, x_0) &= v(\lambda^*(\lambda, x_0), s(\lambda, x_0)) \,.
    \end{split}
\end{equation}
The construction of the (possibly noncompact) double null coordinates is now accomplished. The compactification, see \eqref{comp}, will be discussed in the next part of this Section.

\subsection{Conformal diagram}

\begin{figure*}
    \centering
    \includegraphics[width=1\linewidth]{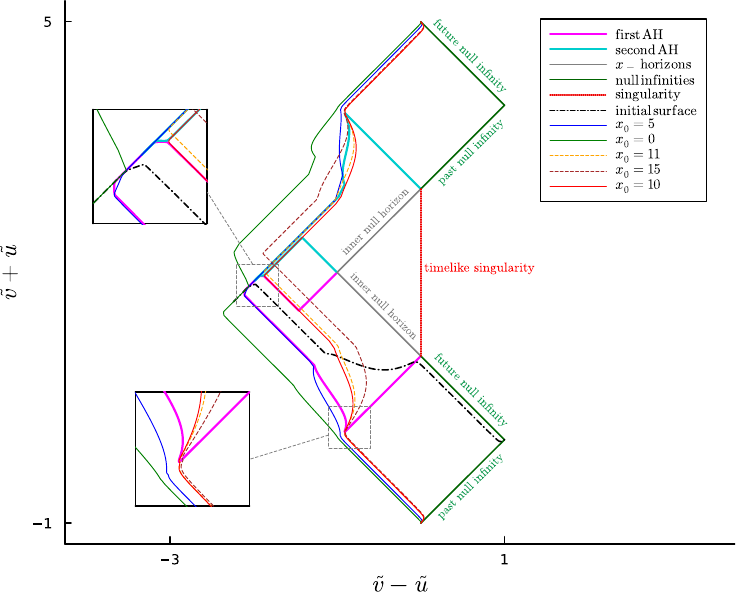}
    \caption{  (color online). The conformal diagram for Oppenheimer-Snyder collapse scenario with the initial profile \eqref{rhoin} and coordinate choices of \eqref{uvcomptrad} and \eqref{in1}. Trajectories of dust particles (radial timelike geodesics) are labeled by $x_0 \in \{0,5,10\}$. The trajectories of test particles (they originate in vacuum at the initial surface $\lambda=T=0$) are labeled by $x_0 \in \{11, 15 \}$. The trajectory of the surface of the dust ball is $x_0 = x_b = 10$. There are two apparent horizons, first AH and second AH, they are null in the vacuum (outside the dust ball), as expected. These null sections of AHs are connected by segments penetrating the dust ball interior. In addition, there is a pair of $x=x_- \approx 2.5718$ horizons, the lower one in the diagram is the Cauchy horizon with respect to the initial surface $T=\lambda=0$. There are two asymptotic regions, each with the pair of past and future null infinities. The future asymptotic region as well as the timelike singularity were plotted thanks to the extension of the coordinate system discussed in Appendix \ref{extension}. The causal structure is precisely the same as in the one displayed in Figure \ref{diag_OS_ones}. We have taken $\rho_0 = 0.007347$ (Planck units). For clarity, Figure~\ref{dia_OS_notraj} in Appendix~\ref{app_add_plots} presents a version of the diagram with fewer plotted objects.
 }
    \label{diag_OS}
\end{figure*}

\begin{figure*}
    \centering
    \includegraphics[width=1\linewidth]{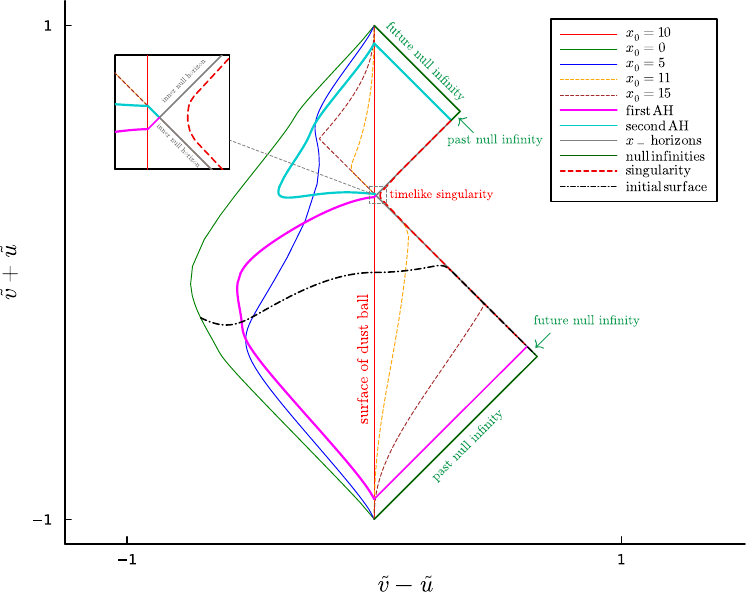}
    \caption{(color online). The conformal diagram for Oppenheimer-Snyder collapse scenario with the initial profile \eqref{rhoin} and coordinate choices of \eqref{uvcompnonstand} and \eqref{in2}. Trajectories of dust particles (radial timelike geodesics) are labeled by $x_0 \in \{0,5,10\}$. The trajectories of test particles (they originate in vacuum at the initial surface $\lambda=T=0$) are labeled by $x_0 \in \{11, 15 \}$. The trajectory of the surface of the dust ball is $x_0 = x_b = 10$. There are two apparent horizons, first AH and second AH, they are null in the vacuum (outside the dust ball), as expected. These null sections of AHs are connected by segments penetrating the dust ball interior. In addition, there is a pair of $x=x_-\approx 2.5718$ horizons, the lower one in the diagram is the Cauchy horizon with respect to the initial surface $T=\lambda=0$. There are two asymptotic regions, each with the pair of past and future null infinities. The future asymptotic region as well as the timelike singularity were plotted thanks to the extension of the coordinate system discussed in Appendix \ref{extension}. The causal structure is precisely the same as in the one displayed in Figure \ref{diag_OS}. We have taken $\rho_0 = 0.007347$ (Planck units). }
    \label{diag_OS_ones}
\end{figure*}

We now numerically compute the conformal diagram for the Oppenheimer-Snyder collapse scenario for two choices of initial data determining the double null coordinates. The results (complete diagrams) will be presented in two diagrams (Figures \ref{diag_OS} and \ref{diag_OS_ones}) whose visual appearance differs from each other, but whose causal structure is exactly the same in both cases. The dissimilarity of diagrams is a consequence of abovementioned two distinct choices of the initial conditions while constructing the double null coordinates --- see discussion around \eqref{in1} and \eqref{in2}. That is, each choice of the initial conditions yields a specific coordinate system. Given a spacetime geometry (metric), we can always switch between coordinate systems—hence, the causal properties remain the same in both diagrams. That is, there exists a diffeomorphism (coordinate transformation) that maps the spacetime in Figure \ref{diag_OS} to that in Figure \ref{diag_OS_ones}, and vice versa.

Let us focus on the 'traditionally' looking diagram in the Figure \ref{diag_OS}. We utilize the double null coordinates \eqref{uvlamx0_f} as constructed in the previous paragraphs with the choice of initial conditions of \eqref{in1}. The coordinates \eqref{uvlamx0_f} are not compact with that choice. Hence, according to the strategy in Section \ref{secStrategy} we propose the following compactification
\begin{equation} \label{uvcomptrad}
        \begin{split}
        \tilde{u}(\lambda, x_0) \rightarrow \pm 1/\pi\arctan u(\lambda, x_0) + c_u \, , \\
        \tilde{v}(\lambda, x_0) \rightarrow \pm 1/\pi \arctan v(\lambda, x_0) + c_v \, ,
    \end{split}
\end{equation}
where $u(\lambda, x_0)$ and $v(\lambda, x_0)$ are given in \eqref{uvlamx0_f}. Here we have diagram blocks' constants $c_u \in \{0,1,2 \}$ and $c_v \in \{0, 1, 2\}$. We will later explain their role (as well as the $\pm$ signs in \eqref{uvcomptrad}).
For the nonstandardly looking diagram in the Figure \ref{diag_OS_ones}, we have a simpler compactification available
\begin{equation} \label{uvcompnonstand}
        \begin{split}
        \tilde{u}(\lambda, x_0) \rightarrow 1/\pi\arctan u(\lambda, x_0)  \, , \\
        \tilde{v}(\lambda, x_0) \rightarrow 1/\pi \arctan v(\lambda, x_0) \, .
    \end{split}
\end{equation}
Indeed, the whole spacetime is covered by \eqref{uvcompnonstand}.

The diagrams in Figures \ref{diag_OS} and \ref{diag_OS_ones} have exactly the same causal structure and display the following objects. The initial surface $T=\lambda=0$, where the chosen initial dust profile \eqref{rhoin} holds, is represented by dash-dot curve. We fix initial location of the surface of dust ball to $x_b =10$. The dust particles' trajectories are labeled by $x_0 \in \{0, 5, 10\}$ --- note that $x_0$ remains constant during their entire time evolution, however, $\lambda \in \{-\infty, \infty \}$ varies on them. Notice that the trajectory $x_0 = 0$ is also the origin of the radial coordinate (the radius of 2-spheres) $x(\lambda, x_0 =0) =0$. We plot also two trajectories of test particles $x_0 \in \{11, 15\}$, that is the timelike geodesics that originate outside the initial dust ball i.e. $x_0>x_b$. Apparently, they penetrate the collapsing dust ball when the dust particles $x_0 \in \{0, 5, 10\}$ already bounced before. 

We distinguish between two families of horizons: i) apparent horizons that can be penetrated (crossed) by timelike geodesics (referred to as the 'first AH' and 'second AH' in Figures~\ref{diag_OS} and~\ref{diag_OS_ones}) and ii) apparent horizons that cannot be penetrated (crossed) by any timelike geodesics. We label these horizons as 'inner null horizons', and the radial coordinate $x = x_- = const.$ remains constant along them.

First, let us discuss the former family. The first apparent horizon ('first AH'), penetrable by timelike geodesics, corresponds to roots of $1-N^x(\lambda, x_0)$, the second one ('second AH') on the other hand, corresponds to roots of $1+N^x(\lambda, x_0)$. Importantly, both horizons are null in the vacuum (outside the dust ball). The first AH crosses the initial surface $T=\lambda=0$ once, it also crosses the surface of the collapsing dust ball $x_0=x_b=10$ twice --- at these crossing points the horizon becomes null. In the interior of the dust ball, the first AH switches between being timelike or spacelike and connects the exterior's null sections. Similar observations apply to the second AH with exception that it does not cross the surface $T=\lambda=0$.

We now turn to a discussion of the latter family, namely the 'inner null horizons', on which $x = x_- = \text{const.}$ Consider future-directed radial null geodesics departing from the trapped segment of the dust ball surface—i.e., the portion enclosed by the first AH (see Figures~\ref{diag_OS} and~\ref{diag_OS_ones}). Along these geodesics, the radial coordinate $x$ decreases as they propagate into the future, as expected due to their trajectory through the trapped region. Eventually, each such geodesic intersects the surface $x = x_- = const.$, which corresponds to a root of $1 - N^x_{\rm OS}(T,x)$, where $N^x_{\rm OS}(T,x)$ is given in~\eqref{shiftOS}. This surface indeed marks the location of an apparent horizon, since apparent horizons are located at spacetime points where the null geodesics' expansion vanishes (see Section \ref{subsecParttraj}). The apparent horizon at $x = x_- = \text{const.}$ (inner null horizon) is null because it lies in the vacuum region. The geodesics then terminate at $x = 0$, which corresponds to the location of the timelike singularity. We note that no timelike geodesic can cross this inner null horizon. The timelike geodesics, labeled by $x_0 \in [0, \infty)$, approach the inner null horizon at $x = x_-$ more closely as $x_0$ increases, yet they never cross it. Indeed, this inner null horizon also serves as a Cauchy horizon.  A similar analysis applies to past-directed geodesics departing from the anti-trapped segment of the dust ball enclosed by the second AH. These geodesics also intersect the surface $x = x_-$ before terminating at $x = 0$.  In summary, Figures~\ref{diag_OS} and~\ref{diag_OS_ones} exhibit two distinct inner null horizons—each connecting the null segments of the first and second AH, respectively.

The computed conformal diagrams indicate that the studied spacetime of quantum-corrected Oppenheimer-Snyder scenario qualitatively resembles Reissner-Nordstr\"om spacetime. The collapsing matter bounces and reemerges in a new universe. Indeed, there are two asymptotic regions enclosed by two pairs of past and future null infinities --- see Figures \ref{diag_OS} and \ref{diag_OS_ones}. Moreover, a timelike singularity is present, where the radius goes to zero $x \rightarrow 0$ and curvature invariants e.g. Kretschmann scalar diverge. Furthermore, the diagrams show that the spacetime of our study represents black-to-white hole transition, where null sections of first AH generate a pair of black hole horizons whereas the null sections of second AH generate a pair of white hole horizons. 
Some aspects of the construction, in particular the role of block constants $c_u$, $c_v$ and $\pm$ signs in the compactified coordinates \eqref{uvcomptrad} used in building the diagram in Figure \ref{diag_OS} are discussed in Appendix \ref{const}.

It is necessary to emphasize, that
some regions in diagrams in Figures \ref{diag_OS} and \ref{diag_OS_ones} were not plotted precisely in coordinates as defined in $\eqref{uvcomptrad}$ or $\eqref{uvcompnonstand}$. Here we recall that the pair $(\lambda, x_0)$ spans a congruence of timelike geodesics representing trajectories of dust (or test) particles originating at the initial surface $T=\lambda=0$. These timelike geodesics never enter both the region enclosed by $x_-$ horizons and the singularity --- see Figures \ref{diag_OS} and \ref{diag_OS_ones} --- and the post-bounce exterior vacuum region located beyond (first) null section of second AH. In addition, note that in Figures \ref{diag_OS} and \ref{diag_OS_ones} the test particle trajectories $x_0\in \{11,15\}$ once enter the dust ball, they never leave it. So, to plot the remaining regions not covered by the pair $(\lambda,x_0)$, we perform an \emph{analytic extension} of our coordinate system as defined in \eqref{uvcomptrad} and \eqref{uvcompnonstand}, see Appendix \ref{extension} for the details. 


\section{Nonhomogeneous Collapse} 
\label{secNH}

With the Oppenheimer-Snyder collapse scenario extensively studied in the previous section, we now turn to the main problem of this work, that is, the investigation of nonhomogeneous dust collapse. We emphasize that the techniques employed for that homogeneous variant, up to the choice of the initial conditions in \eqref{systemSstarU} and \eqref{systemSstarV}, will work also for the studied nonhomogeneous case. We choose the following Gaussian initial nonhomogeneous dust profile for which we will determine the causal structure
\begin{equation} \label{densityNH}
    \rho_\mathrm{NH}(T=\lambda=0, x_0) = \rho_0 \exp\left( \frac{-x_0^2}{2 \mathdutchcal{r}^2} \right) \,,
\end{equation}
where $\mathdutchcal{r}$ is simply a constant representing the standard deviation of the Gaussian. As in the OS collapse scenario, the first objective is to determine the geometry (metric components) via the method of characteristic presented in Section \ref{secMODEL}. With the choice of \eqref{densityNH}, $\beta^\mathrm{in}$ from \eqref{initial} now becomes
\begin{equation} \label{nhbetain}
\begin{split}
            \beta^\mathrm{in}_\mathrm{NH}(x_0) = -\arcsin \Biggl(  \frac{8 \pi G \gamma^2 \Delta \rho_0}{x_0^3} \biggl[ e^{\frac{-x_0^2}{2 \mathdutchcal{r}^2}} \mathdutchcal{r}^2 x_0 + \\ 
            \sqrt{\frac{\pi}{2}} \mathdutchcal{r}^3 \mathrm{erf}\left( \frac{x_0}{\sqrt{2} \mathdutchcal{r}} \right) \biggr]  \Biggr)
\end{split}
\end{equation}
where $\mathrm{erf}$ is the integral of Gaussian distribution, namely $\mathrm{erf}(z) := \frac{2}{\sqrt{\pi}} \int^z_0 e^{-z^2} \mathrm{d}z$. One can plug \eqref{nhbetain} to \eqref{sol1} in order to obtain functions $T(\lambda, x_0)$, $x(\lambda, x_0)$ and $\beta(\lambda, x_0)$ for nonhomogeneous scenario.
Due to the complexity of the problem, we do not extract analytic form of metric components in PG coordinates \eqref{metric2}. Instead one works with the line element \eqref{metricLAMX0} whose metric components are determined via abovementioned functions $T(\lambda, x_0)$, $x(\lambda, x_0)$ and $\beta(\lambda, x_0)$. 

The procedure for finding relevant objects that we will display on the conformal diagram is similar the one presented for Oppenheimer-Snyder collapse scenario discussed in Section \ref{secOS}, because the analysis for e.g. dust particles's trajectories and apparent horizons is general enough. The initial surface is again given by $T=\lambda = 0 $ with $x_0 \in \{0, \infty\}$ as a parameter there. The dust particles trajectories (timelike radial geodesics) are affinely parametrized by $\lambda \in \{-\infty, \infty \}$ and $x_0$ remain constant for each member of the congruence. Furthermore, the first AH again is located at the roots of $1-N^x(\lambda, x_0)$. For now, we postpone the discussion of the second apparent horizon alleged to be located at roots of $1+N^x(T,x)$. We will see in the next paragraphs that this issue requires more cautious approach. 

Before proceeding further however, we first need to address one crucial phenomenon, that seriously limits the range of probing the spacetime structure by the methods applied here -- the so-called \emph{shell-crossing singularities} formation during the dust collapse.

\subsection{Shell-crossing singularities}

The shell crossing singularity is a spacetime location where the energy density \eqref{density} diverges despite the fact that geodesics passing through that location are extendible (see e.g. \cite{Szekeres:1995gy} for a more general definition). Within the model presented in Section \ref{secMODEL}, they were seminally studied in \cite{Fazzini:2023ova}. We will improve on the previous studies in the sense that we will pinpoint where shell-crossing singularities form in spacetime, and further, we will display them on the conformal diagram. 

\begin{figure*}
   \centering
    \includegraphics[width=0.8\linewidth]{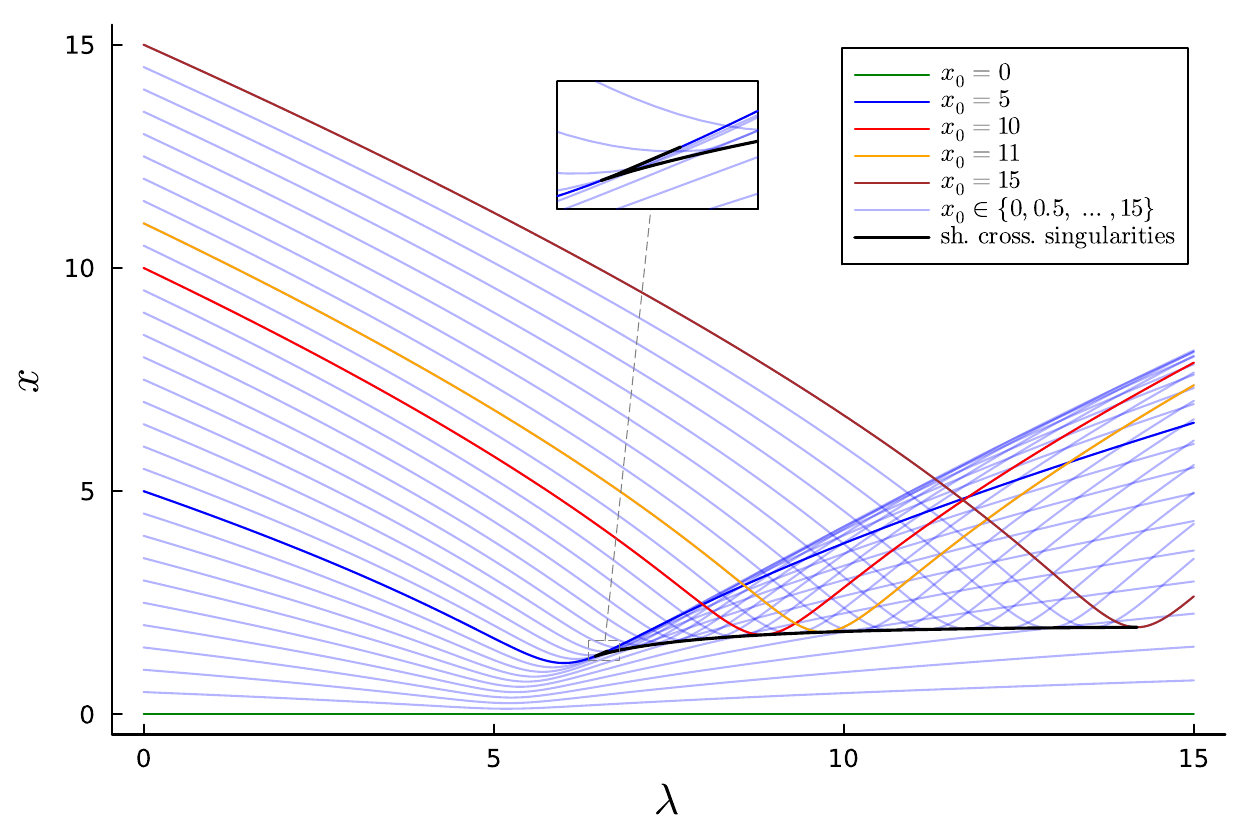}
    \caption{(color online). Timelike geodesics (dust particles' trajectories) and shell-crossing singularities depicted on $x$ (radial coordinate) vs. $\lambda$ (affine parameter) plane for nonhomogeneous dust collapse with the initial profile given in \eqref{densityNH}. Each geodesic correspond to the curve characterized by $x(\lambda, x_0 = const.)$, where $x_0$ labels the respective ones. shell-crossing singularities are plotted for $x_0 \in (x_0^\mathrm{min} \approx 4.893, 15)$. We have taken $\mathdutchcal{r} = 5$, $\rho_0 \approx 0.007347$ (Planck units).  }
    \label{shellcross}
\end{figure*}

To begin with, we aim to find the formula for the energy density merely as a function of the pair $(\lambda, x_0)$. So far, the formula \eqref{density2} is not satisfactory for us --- it contains unknown term proportional to $\partial_x \beta$ when expanded. We determine the desired $\partial_x \beta$ as a function of $(\lambda, x_0)$ as follows. Firstly, we differentiate \eqref{betadot} with respect to $x$ to obtain
\begin{equation} \label{betadiffx}
\begin{split}
  x \left(\frac{\partial \beta}{\partial x}\right)^2\cos (2 \beta)+& 4 \frac{\partial \beta}{\partial x} \sin \beta  \cos \beta
    \\ + x & \frac{\partial^2 \beta}{\partial x^2} \sin \beta  \cos \beta + \gamma  \sqrt{\Delta } \frac{\partial^2 \beta}{\partial x \partial T} = 0 \, .
\end{split}
\end{equation}
Secondly, we define a new function $\sigma(T,x):=\partial_x \beta (T,x)$ so that \eqref{betadiffx} now becomes
\begin{equation} \label{sigmaeq}
    \begin{split}
  x \, \sigma^2\cos (2 \beta)+& 4 \sigma \sin \beta  \cos \beta
    \\ + x & \frac{\partial \sigma}{\partial x} \sin \beta  \cos \beta + \gamma  \sqrt{\Delta } \frac{\partial \sigma}{ \partial T} = 0 \, .
\end{split}
\end{equation}
We will solve the above equation along the characteristic curves discussed in Section \ref{subSeccharacteristicslamx0}. That is, we pick $x_0 = const.$ curve and rewrite $\sigma(x, T) \rightarrow \sigma(\lambda, x_0) $, $\beta(x,T) \rightarrow \beta(\lambda, x_0) $, $x \rightarrow x(\lambda, x_0)$. Thus, \eqref{sigmaeq} together with a choice of initial condition become
\begin{equation} \label{eqsigma}
\begin{split}
    \frac{\mathrm{d}\sigma}{\mathrm{d}\lambda} &= - 2 \sin \left[ 2 \beta(\lambda, x_0)  \right] \sigma(\lambda, x_0) \\  &- x(\lambda, x_0) \cos\left[ 2 \beta(\lambda, x_0) \right] \sigma (\lambda, x_0)^2 \,, \\
   &\quad \;\; \sigma(\lambda = 0) = \frac{\mathrm{d}\beta_\mathrm{NH}^\mathrm{in}}{\mathrm{d}x_0} \,,
\end{split}
\end{equation}
where functions $\beta(\lambda, x_0)$ and $x(\lambda, x_0)$ are given in \eqref{sol1}. One can find analytic solution to \eqref{eqsigma} with the use of \textsc{Mathematica} software. Having the equation for $\sigma$ solved, we rewrite the expression for the energy density $\rho$ given in \eqref{density2} as
\begin{equation}
\begin{split} \label{densityshellcross}
    \rho(\lambda,x_0)=& \frac{3 \sin^2 \left[ \beta(\lambda, x_0)\right]}{8 \pi G \gamma^2 \Delta} \\ &+  \frac{x(\lambda, x_0) \cos \beta(\lambda,x_0) \sin \beta(\lambda, x_0) \sigma(\lambda, x_0)}{4 \pi G \gamma^2 \Delta} \,. 
\end{split}
\end{equation}
Thus, the energy density can become infinite whenever the second term in \eqref{densityshellcross} containing $\sigma$ blows up. It turns out that the divergence can happen in finite $\lambda$-time for the radial timelike geodesics for which $x_0 > x_0^\mathrm{min}$, where $x_0^\mathrm{min}$ is a certain value (keep in mind that $x_0$ is constant on the chosen member of timelike geodesic congruence). In other words, a shell-crossing occurs where for a given pair of coordinates $(\lambda, x_0)$, the function $\sigma(\lambda, x_0)$ is infinite.  In particular, we present results of numerical investigations in Figure \ref{shellcross}. Apparently shell-crossing singularities are forming when 'late-time geodesics' (large initial $x$ at $\lambda =0$) are crossing the early-time geodesics (small initial $x$) that have already bounced and became trajectories of the outgoing dust particles --- see Figure \ref{shellcross}.

\subsection{Conformal diagram} \label{sectionNonhomoConfo}

\begin{figure*}
    \centering
    \includegraphics[width=1\linewidth]{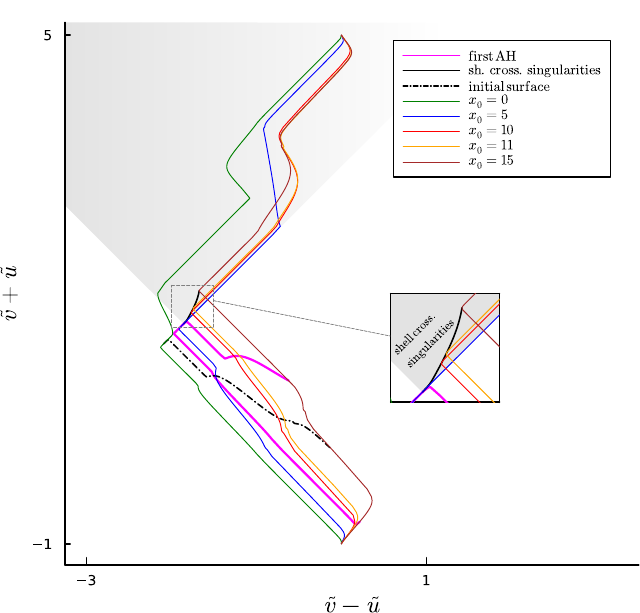}
    \caption{(color online). The conformal diagram of nonhomogeneous dust collapse for the initial profile \eqref{densityNH} and coordinate choices of \eqref{uvcomptrad} and \eqref{NHin1}. We were able to cover the spacetime regions where $ 0\leq x_0 \leq 15 $. Trajectories of infalling dust particles (timelike geodesics) labeled by values of $x_0$ are bouncing and then reexpanding. The first AH is touching the timelike geodesic $x_0 =15 $ at two distinct spacetime points. The shell crossing singularities, plotted for $x_0 \in [x^\mathrm{min}_0 \approx 4.893, 15]$, are being in the vicinity of bouncing late-time timelike geodesics $x_0 \in \{10,11,15\} $. The shadow of the shell-crossing singularities (grayed-out area) yield a region of spacetime for which the results are no longer reliable. The causal structure is compatible with the one displayed in Figure \ref{diag_nonhomo_ones}. We have taken $\mathdutchcal{r} = 5$, $\rho_0 \approx 0.007347$, $x^\mathrm{NH}_{u} =7.5$ and $x^\mathrm{NH}_{v}=15$  (Planck units).   }
    \label{diag_nonhomo}
\end{figure*}

\begin{figure*}
    \centering
    \includegraphics[width=1\linewidth]{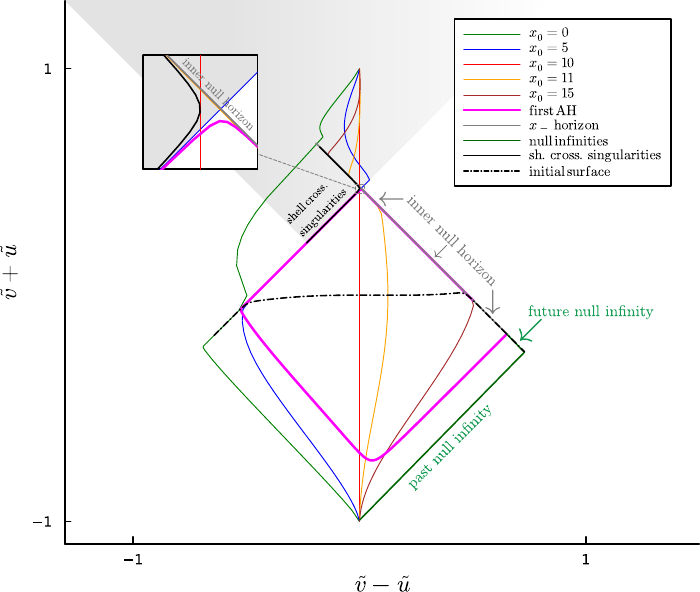}
    \caption{(color online). The conformal diagram for nonhomogeneous dust collapse for the initial dust profile \eqref{densityNH} and coordinate choices of \eqref{uvcompnonstand} and \eqref{NHin2}. Trajectories of infalling dust particles (timelike geodesics) labeled by values of $x_0$ are bouncing and then reexpanding. The first AH is originating at the crossing point of the inner null horizon and the future null infinity. Subsequently, the first AH is terminating at the point located at the inner null horizon $x=x_-$. The radial coordinate is constant on the entire inner null horizon, that is, $x=x_- \approx 2.0195$. The shell crossing singularities, plotted for $x_0 \in [x^\mathrm{min}_0 \approx 4.893, 15]$, are being formed in the vicinity of bouncing late-time timelike geodesics $x_0 \in \{10,11,15\} $. The shadow of the shell-crossing singularities (grayed-out area) yield a region of spacetime for which the results are no longer reliable. For the overlapping portions of spacetime, the causal structure is compatible with the one displayed in Figure \ref{diag_nonhomo}, however, here we were able to probe much more when compared to that companion diagram. In particular, we see that the future null infinity evolves into inner null horizon. Hence, the collapse cannot take place within single asymptotic region. An extension of spacetime beyond inner null horizon (Cauchy horizon) is needed. We have taken $\mathdutchcal{r} = 5$, $\rho_0 \approx 0.007347$ and $x^\mathrm{NH}_{u}= x^\mathrm{NH}_{v} =10$. (Planck units). For clarity, Figure~\ref{dia_nonhomo_noappa} in Appendix~\ref{app_add_plots} presents a version of the diagram with fewer plotted objects.}
    \label{diag_nonhomo_ones}
\end{figure*}

With the role of the shell-crossing singularities understood, we now can build and discuss conformal diagrams for the nonhomogeneous dust collapse. These diagrams are presented in Figures \ref{diag_nonhomo} and \ref{diag_nonhomo_ones}. We emphasize that the diagram in Figure \ref{diag_nonhomo} presents less spacetime regions than the one in Figure \ref{diag_nonhomo_ones}. This is a consequence of the complexity of numerical calculations needed to extract the former one. In other words, not every spacetime region depicted in the Figure \ref{diag_nonhomo_ones} is also depicted in the Figure \ref{diag_nonhomo}. In the following paragraphs we will explain this issue in more detail.

To plot the diagram in Figure \ref{diag_nonhomo} we employed the compactified coordinates \eqref{uvcomptrad} with the choice of the following initial conditions in order to solve the systems \eqref{systemSstarU} and \eqref{systemSstarV}
\begin{equation} \label{NHin1}
\begin{split}
x_{\mathrm{in}, s=0}^u \rightarrow x^\mathrm{NH}_{u}\,,\;\; f^u_{\mathrm{in}, s=0}& \rightarrow -1/[1-N^x(\lambda = \lambda^*, x_0 = x^\mathrm{NH}_{u}) ]\,, \\
x_{\mathrm{in}, s=0}^v \rightarrow x^\mathrm{NH}_{v} \,, \;\; f^v_{\mathrm{in}, s=0} &\rightarrow 1/[1+N^x(\lambda = \lambda^*, x_0 = x^\mathrm{NH}_{v}) ] \,,
\end{split}
\end{equation}
where $x^\mathrm{NH}_{u}$, $x^\mathrm{NH}_{v}$ are constants and their chosen values are listed in caption of Figure \ref{diag_nonhomo}. In the process of solving the system \eqref{systemSstarV} (with the initial conditions \eqref{NHin1}) and subsequently integrating the solution in order to determine the relation \eqref{uvlamx0_f} some difficulties arose when probing the spacetime regions for $x_0>x^\mathrm{NH}_{v}$ due to computational complexity. For that reason, we only managed to depict in Figure \ref{diag_nonhomo} those spacetime regions that are enclosed by the geodesics $x_0 = 0$ and $x_0 = x^\mathrm{NH}_{v}$ (see the diagram). It is worth noting however that this problem is resolved with the different choice of initial conditions than \eqref{NHin1}, see next paragraphs discussing the diagram in Figure \ref{diag_nonhomo_ones}. Nonetheless, in the diagram in Figure \ref{diag_nonhomo} we plot the dust particle trajectories for $x_0 \in \{0, 5, 10, 11, 15 \}$. The first AH is crossing the initial surface $T=\lambda=0$ and touching the geodesic $x_0 = 15$ at two distinct points. 

Next, we note that the picture of formation of shell crossing singularities in the diagram in Figure \ref{diag_nonhomo} is consistent with previously discussed one in Figure \ref{shellcross}. That is, in Figure \ref{diag_nonhomo} we can see that the geodesics $x_0 \in \{10,11,15\}$ are entering (before bouncing) the region where geodesics $x_0 \in \{0, 5\}$ are already outgoing after the bounce in their past. The shell-crossing singularities form close to the bouncing points of the late-time geodesics $x_0 \in \{10,11,15\}$. 

At this point some comments ought to be made with respect to the second AH. While the shell-crossing singularities are the spacetime points where the energy density \eqref{densityshellcross} becomes infinite, one could in principle attempt constructing an extension of the geodesics past them. However, past them the dynamical equations presented in Section \ref{secMODEL} are no longer reliable, as they do not account for the ``transversal'' flux of matter carried along the geodesics being crossed. For that reason, future light cones originating from the points of shell-crossing singularities yield a 'shadow' displayed in Figure \ref{diag_nonhomo} as the grayed-out area --- this is the spacetime region where the geometry as determined by the model of our consideration is no longer physically reliable. For that reason, we believe that that there is no sense in plotting the second AH (resulting from naive computation of the roots of $1+N^x(\lambda, x_0)$) that would be located in the shadow.

The diagram in Figure \ref{diag_nonhomo_ones} presents a causal structure compatible with the one in Figure \ref{diag_nonhomo}. It is based on the following choice of the initial conditions for constructing the double null coordinates
\begin{equation} \label{NHin2}
\begin{split}
x_{\mathrm{in}, s=0}^u \rightarrow x^\mathrm{NH}_{u}\,,\;\; f^u_{\mathrm{in}, s=0}& \rightarrow -1\,, \\
x_{\mathrm{in}, s=0}^v \rightarrow x^\mathrm{NH}_{v} \,, \;\; f^v_{\mathrm{in}, s=0} &\rightarrow 1 \,,
\end{split}
\end{equation}
where the choices of constants $x^\mathrm{NH}_{u}$ and $x^\mathrm{NH}_{v}$ are listed in the caption in Figure \ref{diag_nonhomo_ones}. Similarly to the previous case, these initial conditions determine the functions $f^u,f^v$ as solutions to \eqref{systemSstarU}, \eqref{systemSstarV} and, subsequently, they allow us to employ the compactified coordinate system \eqref{uvcompnonstand} in order to construct the diagram. The advantage of these coordinates for nonhomogeneous case is the fact that, in opposite to the coordinates employed in the previous paragraph, there are no computational difficulties in exploring the spacetime beyond the limiting timelike geodesic $x_0=15$. Indeed, the dust particle trajectories (timelike goedesics) behave qualitatively the same, shell-crossing singularities are also similarly located, however, we were able to explore the spacetime up to an asymptotic infinity connected to the initial slice $T=\lambda =0$. In Figure \ref{diag_nonhomo_ones} we can see that there are two (future and past) null infinities.

Similarly to the OS collapse scenario, alongside the first AH --- an apparent horizon that can be penetrated (crossed) by timelike geodesics --- there also exists an apparent horizon that cannot be crossed by any timelike geodesic. We again refer to this latter apparent horizon as the 'inner null horizon', on which the radial coordinate remains constant, $x = x_- = \text{const.}$. Indeed, the inner null horizon is also an apparent horizon, because it corresponds to spacetime points where $1-N^x \to 0$.

The first AH starts and ends in two distinct points that are connected by $x_-$ horizon similarly as in the Oppenheimer-Snyder collapse scenario discussed in Section \ref{secOS}. The inner null horizon $x=x_-=const.$ evolves from the future null infinity, subsequently penetrating the trajectories of dust particles and is extending up to the origin of the radial coordinate (equivalently $x_0=0$ geodesic). However, we again note that the shadow of shell-crossing singularities yields a spacetime region that cannot be reliably probed via applied methods (due to shell-crossing singularities), so $x_-$ is in that region a mere conjecture.

At this point some elaboration is needed as to how the inner null horizon $x = x_- = \text{const.}$ was determined for the nonhomogeneous collapse scenario under consideration. Unlike the OS collapse scenario, we do not have an analytic form of $N^x(T,x)$ at our disposal, however, we can still perform a similar, though now numerical, analysis. Consider future-oriented congruence of null geodesics departing from the 'straight-line' timelike geodesic in Figure~\ref{diag_nonhomo_ones}. Let $x_0$ be a parameter along these geodesics. Solving the radial null condition $\mathrm{d}\mathdutchcal{s}^2 = 0$, derived from equation~\eqref{metricLAMX0}, yields a function $\lambda(x_0)$ along these geodesics. More precisely, one has to solve the differential equation $
\frac{\mathrm{d}\lambda }{\mathrm{d}x_0} = \frac{1}{\gamma \sqrt{\Delta}} \frac{\partial x}{\partial x_0} $ with the initial condition $\lambda(x_{0_{\rm sl}}) = \lambda_{{\rm sl}}$. For the 'straight-line' timelike geodesic in Figure \eqref{diag_nonhomo_ones} we have $x_{0_{\rm sl}} =10$  and $\lambda_{{\rm sl}}$ parametrizes it. Once this step is accomplished, we evaluate the functions $x(\lambda(x_0),x_0)$, $1 - N^x(\lambda(x_0),x_0)$, and $v(\lambda(x_0),x_0)$, where $v$ is to-be-compactified null coordinate defined in equations~\eqref{uvlamx0_f} and~\eqref{NHin2}.

\begin{figure}
    \centering
    \includegraphics[width=1\linewidth]{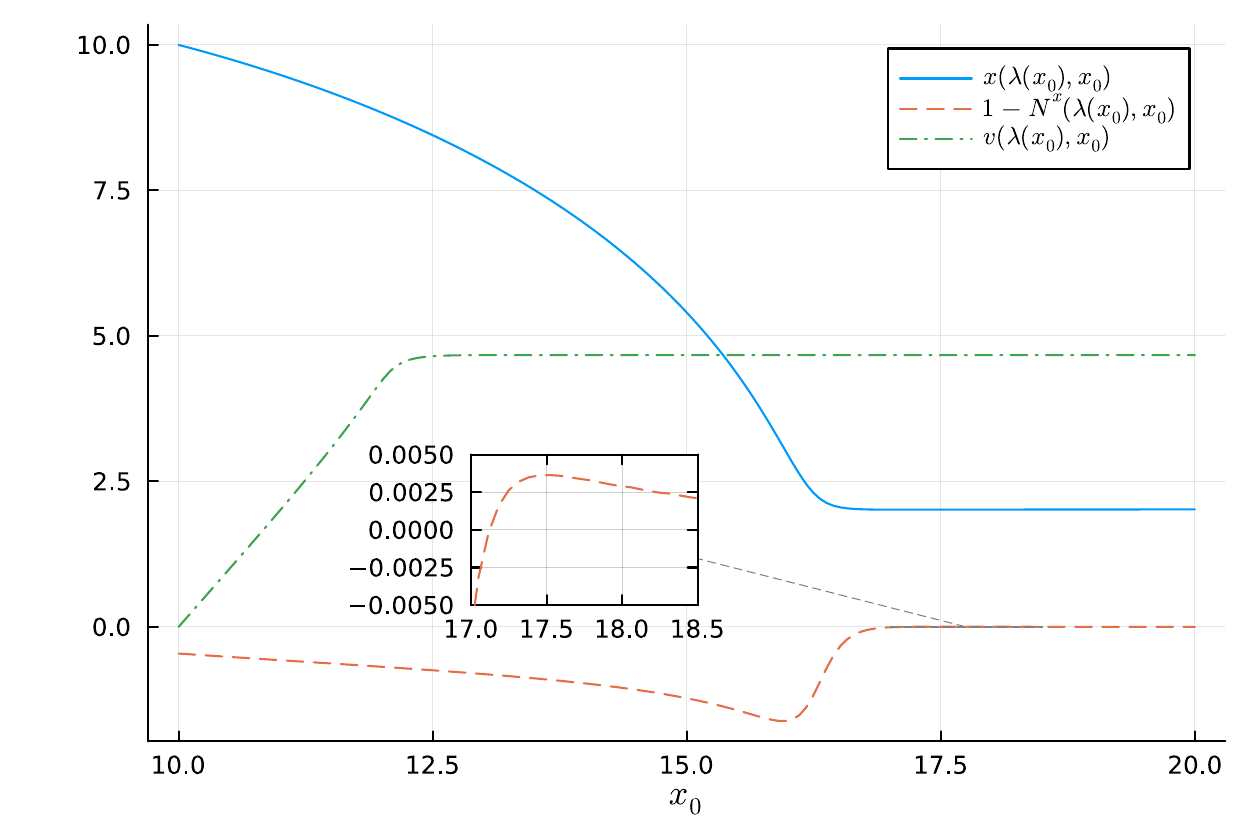}
    \caption{(color online). The functions $x(\lambda(x_0), x_0)$, $1 - N^x(\lambda(x_0), x_0)$, and $v(\lambda(x_0), x_0)$, evaluated along a single null geodesic departing from the 'straight-line' timelike geodesic shown in Figure~\ref{diag_nonhomo_ones}, exhibit asymptotic behavior as the geodesic approaches the inner null horizon in the limit $x_0 \to \infty$. In this limit, each function tends to a constant value, specifically, $x \to x_- \approx 2.0195$, $1 - N^x(\lambda(x_0), x_0) \to 0$, and $v \to v_+ \approx 4.67$. These asymptotic values are valid for all geodesics labeled by $\lambda_{\rm sl} > \lambda_{\rm sl_1} \approx -33.94$. Note also that $1 - N^x(\lambda(x_0), x_0)$ changes sign before the inner null horizon is reached — specifically, when the first AH in Figure~\ref{diag_nonhomo_ones} is crossed.  For this analysis, we have taken $x_{0_{\rm sl}}=10$, $\lambda_{\rm sl} =0$ ,$\mathdutchcal{r} = 5$, $\rho_0 \approx 0.007347$, and $x^\mathrm{NH}_{u} = x^\mathrm{NH}_{v} = 10$. (Planck units).}
    \label{nullgeo}
\end{figure}

We present the results of the above procedure in Figure~\ref{nullgeo} for a single null geodesic belonging to the congruence discussed earlier, which departs from the 'straight-line' timelike geodesic in the region enclosed by the first apparent horizon (AH) in Figure~\ref{diag_nonhomo_ones}. In the figure, the corresponding functions are plotted, and we observe the following: $(i)$ the radial coordinate approaches a limit $x \to x_- = \text{const.}$, $(ii)$ the to-be-compactified coordinate approaches $v \to v^+ = \text{const.}$, and $(iii)$ the quantity $1 - N^x(\lambda(x_0), x_0)$ changes sign once (corresponding to the crossing of the first AH) and then asymptotically approaches zero. These limits hold for every radial null geodesic that departs from the 'straight-line' timelike geodesic for initial parameter values $\lambda_{\rm sl} > \lambda_{\rm sl_1} = \text{const.}$. From this analysis, we infer that every geodesic labeled by $\lambda_{\rm sl} > \lambda_{\rm sl_1}$ approaches a null surface where the radial coordinate is constant and $1 - N^x(\lambda(x_0), x_0)$ vanishes. This implies that the surface that these geodesics approach in the limit $x_0 \to \infty$ is a non-expanding null tube --- an apparent horizon. 

For the remaining null geodesics, labeled by $\lambda_{\rm sl} < \lambda_{\rm sl_1}$, the behavior is qualitatively different. We find that in the limit $x_0 \to \infty$ $(i)$ $x \to \infty$, $(ii)$ the to-be-compactified coordinate approaches $v \to v^+ = \text{const.}$, and $(iii)$ $1 - N^x(\lambda(x_0), x_0)$ tends to 1. From that we conclude, that these geodesics terminate at future null infinity.

\subsection{Is there a timelike singularity?}

Within the Oppenheimer-Snyder collapse scenario we have observed that there exist the timelike singularity, see Figures \ref{diag_OS} and \ref{diag_OS_ones}. Its existence is not directly resulting from the dynamical equations formulated in Section \ref{secOS}, rather, the singularity comes from the extension of the spacetime, beyond the Cauchy horizon, thanks to the analytic form of the metric component \eqref{schwarz}. From the diagram for the nonhomogeneous collapse in Figure \ref{diag_nonhomo_ones} we can learn that the dynamical equations can determine the spacetime up to the $x_-$ horizon. This is because $x_0 \rightarrow \infty$ there meaning that no characterstic curves (or timelike geodesic) can explore the spacetime beyond $x_-$. In other words, the horizon $x_-$ is a Cauchy horizon. Thus, a natural question arises whether we can probe the spacetime beyond $x_-$. Is there a timelike singularity?

To address the question we will propose the following method for exploring spacetime beyond $x_-$ horizon. We will study a behaviour of the metric components on a radial null geodesic that is perpendicular (on the conformal diagram) to the $x=x_-$ horizon in order to find a possible spacetime (analytic) extension beyond the horizon. In particular, we will study how the function $1-\left(N^x \right)^2$ varies with respect to the affine parameter on the chosen null geodesic and we will look for the possible extensions. In fact, we will build upon the analysis of null geodesics presented in the final paragraphs of Section \ref{sectionNonhomoConfo}, extending it to affine parametrization of null geodesics that cross the inner null horizon.

We work with coordinates $(\lambda, x_0)$ and the corresponding line element \eqref{metricLAMX0}. Let $x_0$ be a parameter on a radial geodesic that will eventually cross the horizon $x=x_-$. Solving the the null condition $\mathrm{d}\mathdutchcal{s}^2 =0$ allows us to write the tangent vector to the null geodesic as
\begin{equation}
    n^\alpha  = \left( \frac{1}{\gamma \sqrt{\Delta}} \frac{\partial x}{\partial x_0}, 1 \right) \,,
\end{equation}
where we have dropped angular components. The tangent vector satisfies
\begin{equation} \label{nvec}
    n^\alpha \nabla_\alpha n^\beta = \kappa_\mathrm{null} n^\beta \,, \quad \kappa_\mathrm{null}= \frac{\frac{\partial^2 x}{ \partial x_0^2}}{\frac{\partial x}{\partial x_0}} + \frac{2}{\gamma \sqrt{\Delta} } \frac{\partial^2 x}{\partial \lambda \partial x_0} \,.
\end{equation}
Thus, the parameter $x_0$ is not affine. In order to find the affine parametrization we exploit similar techniques as in Subsection \ref{subsecParttraj}: we consider the differential condition for possible affine parameter (here denoted as $p$)
\begin{equation}
    \frac{\mathrm{d}^2p}{\mathrm{d}x_0^2} = \kappa_\mathrm{null} \frac{\mathrm{d}p}{\mathrm{d}x_0} \,,
\end{equation}
for which we pick the following initial conditions
\begin{equation} \label{px0}
    \frac{\mathrm{d}p}{\mathrm{d} x_0} \Bigg|_{x_0=x_0^\mathrm{in,null} } = x_0^\mathrm{in,null} \, \quad p(x_0^\mathrm{in,null})= x_0^\mathrm{in,null} \,,
\end{equation}
where $x_0^\mathrm{in,null}$ is a constant. 

The last ingredient needed to write the desired function $1-\left(N^x(p) \right)^2$ is to find the solution to the following differential equation (together with an initial condition) relating $\lambda$ and $x_0$ at the null geodesic
\begin{equation} \label{nulllamx02}
    \frac{\mathrm{d} \lambda}{ \mathrm{d} x_0} =  \frac{1}{\gamma \sqrt{\Delta}} \frac{\partial x}{\partial x_0} \,. \quad \lambda(x_0^\mathrm{in,null}) = \lambda^\mathrm{in,null} \,.
\end{equation}
Note that the above equation is coming from the null condition $\mathrm{d}\mathdutchcal{s}^2 =0$ and was used to write the $\lambda$-component in \eqref{nvec}. We solve \eqref{px0} and \eqref{nulllamx02} numerically. Finally, we can write
\begin{equation} \label{FFaff}
    1-\left(N^x(p) \right)^2 := 1-N^x(\lambda(x_0(p))), x_0(p))^2 \,,
\end{equation}
where $x_0(p)$ is the inverse of the solution to \eqref{px0} and $\lambda(x_0)$ is the solution to \eqref{nulllamx02}. 

\begin{figure*}
    \centering
    \includegraphics[width=0.7\linewidth]{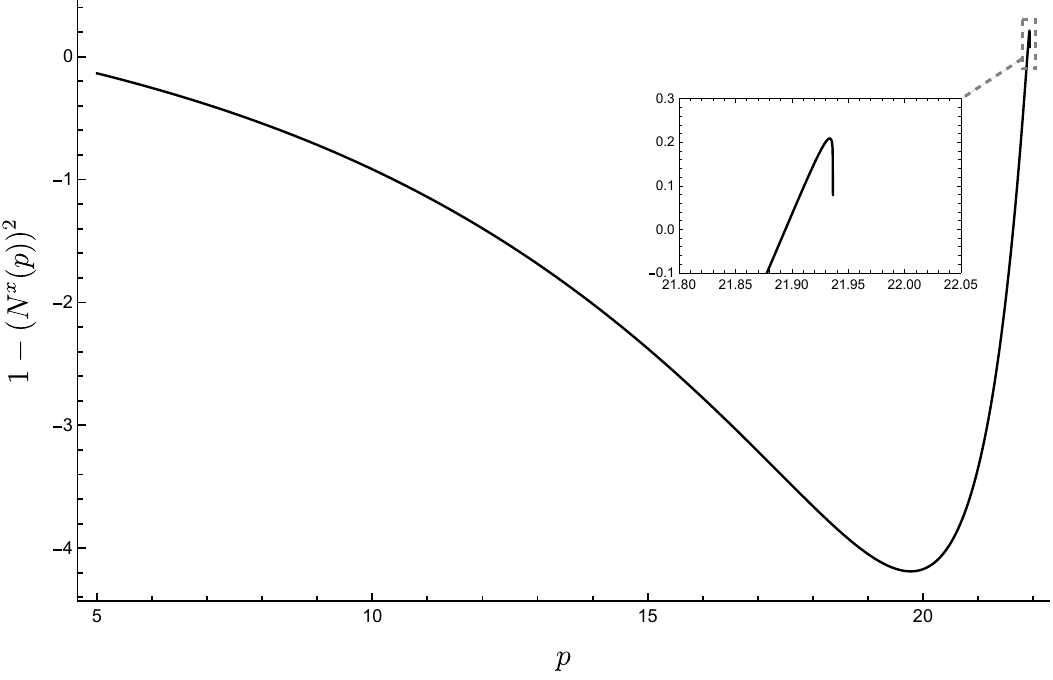}
    \caption{The behaviour of the function $1-(N^x(p))^2$ on the radial null geodesic that starts at the initial surface at $x_0^\mathrm{in,null}=5$ ($p=5$) and then arrives to the vicinity of the Cauchy horizon $x=x_-$. The parameter $p$ is affine. The slope of the function becomes very steep as it get closer to the horizon. For that reason, the function might not be analytically extendable beyond that horizon. We have taken $\mathdutchcal{r} = 5$, $\rho_0 \approx 0.007347$ (Planck units). }
    \label{FFaffine}
\end{figure*}

We present the result of the computation of \eqref{FFaff} in Figure \ref{FFaffine} where we picked the null geodesic that starts at the initial surface at $x_0 = 5$, crosses the first AH and finally arrives to the vicinity of the Cauchy horizon $x=x_-$. Note that the fact that first AH is crossed is reflected by the sign change of $1-\left(N^x(p) \right)^2$. Essentially, the slope of the function becomes very steep as we are getting closer to the Cauchy horizon (large $x_0$). This fact indicates that the function $1-\left(N^x(p) \right)^2$ might not be analytically extendable beyond the horizon (since the derivatives of $1-\left(N^x(p) \right)^2$ tend to infinity). For that reason, we cannot probe the spacetime beyond $x=x_-$ with the methods we have employed. Therefore, the fate of the timelike singularity in the nonhomogeneous collapse scenario remains unknown. 

\section{Discussion}
Throughout this work we have studied dust collapse scenarios within effective loop quantum gravity model that was originally presented in \cite{Kelly:2020lec, Kelly:2020uwj, Husain:2021ojz, Husain:2022gwp, Fazzini:2023ova}. We have focused on rigorous analysis of the causal structure for both the homogeneous Oppenheimer-Snyder collapse scenario and nonhomogeneous collapse scenario. We partly followed the techniques presented in previous works e.g. the method of characteristic equations for partial differential equations, however, we refined them so that we were able to both significantly explore new features of the model and reexamine previous statements concerning the physical aspects of the model.

Meanwhile, in Section \ref{secStrategy} we have developed a general strategy for constructing double null coordinates for a large class of spherically symmetric spacetimes. Remarkably, in later parts of this work, the strategy allowed us to construct the conformal diagrams for nonhomogeneous dust collapse within single coordinate charts. The presented methods can be applied beyond the model of our consideration. Indeed, they serve as a precursor for a general algorithm for the conformal diagram construction in arbitrary spacetime. The key observation that was made is the fact that for the double coordinates their differentials must be exact (vanishing of the exterior derivative) --- then there are no difficulties in defining the appropriate integrals of the differentials.

Our studies of the Oppenheimer-Snyder collapse scenario (presented in Section \ref{secOS}) are summarized through two conformal diagrams presented in Figures \ref{diag_OS} and \ref{diag_OS_ones}. They represent black-to-white hole transition, where the collapsing matter bounces and crosses 2 pairs of horizons, black hole ones and white hole ones. There are two asymptotic regions with pairs of null infinities. The geometry and causal structure we obtained agree with earlier studies \cite{BobulaLOOPS, Bobula:2023kbo, Lewandowski:2022zce, Fazzini:2023scu, Han:2023wxg}. Specifically in the works \cite{BobulaLOOPS, Bobula:2023kbo, Lewandowski:2022zce}, the exterior geometry was determined using the Israel-Darmois junction conditions, by matching it to an interior cosmological geometry governed by effective Loop Quantum Cosmology. In contrast, the approach introduced in \cite{Husain:2021ojz, Husain:2022gwp} and studied in this work is fully dynamical midisuperspace one. Like in the previous works \cite{BobulaLOOPS, Bobula:2023kbo, Lewandowski:2022zce}, we also found a timelike singularity that connects the asymptotic regions. However, in the fully dynamical approach, this singularity does not arise directly from the model’s dynamical equations. Instead, it appears as a result of analytically extending the metric components compatible with the vacuum sector of the model. The model's dynamics can only directly probe the spacetime regions spanned by timelike geodesics originating from initial time slice $T=\lambda=0$ and these regions are free from singularities (other than shell-crossing ones). In fact, the timelike singularity is hidden beyond the Cauchy horizon, and to probe the region beyond that horizon, we analytically extended the spacetime\footnote{We extended the coordinate system as explained in Appendix \ref{extension} and later we utilized the analytic form of extracted metric components in \eqref{schwarz}.}. We emphasize that if the spacetime is extended in a way other than analytically, the timelike singularity may not appear at all. It is also important to note that such an analytic extension was not possible in the more realistic, nonhomogeneous scenario of the model, as we will discuss this issue in the next paragraphs. Additionally, the backreaction from Hawking radiation was ignored in our analysis. Including this effect could significantly change the resulting causal structure.

It is also worth noting, that the causal structure for Oppenheimer-Snyder collapse scenario within the model of our consideration was also independently studied in \cite{Munch:2021oqn}, however, we have not found superluminal dust trajectories or any discontinuities which were found in that reference. In our analysis, the tangent to the dust trajectories is always timelike as it satisfies $l^\alpha l_\alpha =-1$ both in the homogeneous and nonhomogeneous collapse scenarios (see sections \ref{subsecParttraj} and \ref{secNH}). Concerning the discontinuities, also called as 'shock waves', they were also suggested to be present in original works \cite{Kelly:2020lec, Kelly:2020uwj, Husain:2021ojz, Husain:2022gwp}. The idea was that the fact that since the characteristic curves (discussed in Section \ref{secMODEL} in this work) cross with each other they give rise to spacetime discontinuities. In this work, within the OS collapse scenario, we have shown that characteristic curves are timelike geodesics and they can cross with each other while maintaining the spacetime continuity. However, for the studied nonhomogeneous dust collapse (see the next paragraph), the existence of shell-crossing singularities might be interpreted as the symptom of the presence of discontinuities discussed in previous works.

The main objective of this work was to study the causal structure of a nonhomogeneous dust collapse (Section \ref{secNH}). In diagrams in Figures \ref{diag_nonhomo} and \ref{diag_nonhomo_ones} we saw that in the vicinity of the initial surface the collapsing matter (for the chosen initial dust profile) behaves similarly to the case of Oppenheimer-Snyder collapse scenario. However, the relevant differences became visible in the region when dust particles are bouncing. There, the timelike geodesics are crossing with each other and as a result, the shell crossing singularities are formed --- there are points in the diagrams (spacetime) where the energy density becomes infinite. Apparently, the shell-crossing singularities form where late-time timelike geodesic penetrate the early-time ones (the late time ones bounce considerably later than the early ones). Due to the divergences in energy density at the shell-crossing singularities, we conclude that the causal shadow of these singularities marks a region of spacetime that cannot be probed using the dynamical equations discussed in Section \ref{secMODEL}. Perhaps one should work with different gauge choices (coordinates for dynamical equations) than those employed that is Painlev\'e-Gullstrand and areal gauges. 

Furthermore, we attempted to explore the possibility of the existence (or absence) of the timelike singularity in this scenario (keep in mind that the singularity was present in Oppenheimer-Snyder dust collapse scenario). We took a null geodesic that is perpendicular (on the conformal diagram) to the Cauchy horizon and tested whether metric components (living at that geodesic) could be analytically extended beyond that horizon. As a result, we could not find the spacetime extension because the slope of the to-be-extended metric function component became very steep indicating that its derivatives may actually diverge and, consequently, the violation of the analyticity. For that reason, we were unable to extend the spacetime with the employed techniques, thus, the problem of the existence of the singularity for the nonhomogeneous collapse scenario remains open. 

What is remarkable, the conformal diagram in Figure \ref{diag_nonhomo_ones} clearly implies that the nonhomogeneous collapse, for the chosen initial Gaussian configuration, cannot occur within a single asymptotic region. The future null infinity evolves into an inner null horizon. Consequently, alleged black hole explosions or gravitational discontinuities --- potentially signaled by the presence of shell-crossing singularities --- cannot be observed by an external observer residing near future null infinity without crossing a horizon, as indicated by the diagram in Figure \ref{diag_nonhomo_ones}. This contrasts with the conformal diagram presented in \cite{Husain:2021ojz, Husain:2022gwp}. Notably, we were unable to find any initial dust configuration that roughly approximates an astrophysical collapse scenario in which the physical picture we obtained would be altered.

Finally, we note that the possible direction for future investigations could be a formulation of a model accounting for the back-reaction with Hawking radiation. Such an improvement would not only provide more accurate model of black hole evolution, but also could heal the existing (not accounting for backreaction) worrisome features of the model, especially the one concerning existence of the timelike singularity in the homogeneous collapse scenario. With the tools introduced in Section \ref{secStrategy} there should be no significant difficulty with rigorously probing the causal structure for such a model. What is more, some recent (general) studies indicate that the backreaction could also resolve inner horizon instability \cite{Bonanno:2022jjp} -- the issue that is also plaguing the model of our consideration and also many other literature models incorporating quantum aspects of gravity.

\section*{Acknowledgements}
This work was supported in part by the Polish National Center for Science (Narodowe Centrum Nauki – NCN) grant OPUS 2020/37/B/ST2/03604.

\appendix

\onecolumngrid

\section{The role role of blocks' constants $c_u$ and $c_v$} 
\label{const}

Here we explain the role of the constants $c_u$ and $c_v$ as well as the $\pm$ signs in compactified double null coordinates defined in \eqref{uvcomptrad}.

For the sake of an example, consider the trajectory of the surface of the dust ball $x_0=x_b=10$ plotted in Figure \ref{diag_OS}. In the past asymptotic region, we choose $c_u= c_v \rightarrow 0$ and plus sings in \eqref{uvcomptrad}. Later, when the surface reaches the first AH the noncompact $u(\lambda, x_0)$ as defined in \eqref{uvlamx0_f} diverge --- for that reason, to keep the monotonicity of the coordinate we switch the sign to minus (for $\tilde{u}$) and pick $c_u\rightarrow 1$ in the region between null sections of first AH. The pattern is the following --- one has to change both the sign and constant for $\tilde{u}$ or $\tilde{v}$ whenever noncompact $u(\lambda, x_0)$ or $v(\lambda, x_0)$ respectively diverge on the boundary between blocks. Thus, in a (let us say) bouncing region, where every trajectory $x_0 \in \{0,5,10,11,15 \}$ apparently bounces in Figure \ref{diag_OS}, we again switch the sign to plus for $\tilde{u}$ and pick $c_u \rightarrow 2$, $\tilde{v}$ remains unchanged with plus sign and $c_v \rightarrow 0$ there. Next, in the region between null sections of second AH, we pick $c_v \rightarrow 1$ and minus sign for $\tilde{v}$, subsequently, we keep $c_u \rightarrow 2$ and the plus sign for $\tilde{u}$. Lastly, at the discussed trajectory $x_0 =10$ in the upper asymptotic region we have $c_v \rightarrow 2$ and plus sign for $\tilde{v}$ and $c_u \rightarrow 2$ with also the plus sign for $\tilde{u}$.

\section{Extension of coordinate system} \label{extension}

To extend the coordinates \eqref{uvcomptrad} and \eqref{uvcompnonstand} beyond the regions that are spanned by the pair $(\lambda, x_0)$ e.g. vicinity of the timelike singularity and future asymptotic region (these regions are in the vacuum enclosed by the surface of the dust ball located at $x_0=x_b$, see Figure \ref{diag_OS} or Figure \ref{diag_OS_ones}), we introduce the following auxiliary coordinates
\begin{equation}
    \begin{split}
        u^\mathrm{aux}(t, x) = t - \int_0^x \frac{1}{1-(N^x)^2} \mathrm{d}x \,, \\
        v^\mathrm{aux}(t, x) = t + \int_0^x \frac{1}{1-(N^x)^2} \mathrm{d}x \,. 
    \end{split}
\end{equation}
Then, the extension of \eqref{uvcompnonstand} is given by
\begin{equation} \label{extended}
    \begin{split}
       \tilde{u}(t, x) \rightarrow 1/\pi\arctan u(\lambda = \frac{1}{\gamma \sqrt{\Delta}}T[u^\mathrm{aux}(t,x)], x_0=x_b) \,, \\
          \tilde{v}(t, x) \rightarrow 1/\pi\arctan v(\lambda = \frac{1}{\gamma \sqrt{\Delta}}T[v^\mathrm{aux}(t,x)], x_0=x_b) \,,
    \end{split}
\end{equation}
where functions $T[u]$ and $T[v]$ are defined to be the inverses of
\begin{equation}
    \begin{split}
        u(T) &\rightarrow u(\lambda=\frac{1}{\gamma \sqrt{\Delta}}T, x_0=x_b) \,, \\
        v(T) &\rightarrow v(\lambda=\frac{1}{\gamma \sqrt{\Delta}}T, x_0=x_b) \,,
    \end{split}
\end{equation}
where $u(\lambda, x_0)$ and $v(\lambda, x_0)$ are given in \eqref{uvlamx0_f} with the choice of initial conditions \eqref{in2}. With the extended coordinates \eqref{extended} we can plot the timelike singularity for which we have $x=0$ and $t\in\{-\infty, \infty\}$. In addition, we plot the null infinities in the future asymptotic region --- we have $x\rightarrow \infty$ and $t\in \{ -\infty, \infty \}$.

The procedure of extending coordinates presented in the previous paragraph is valid for the coordinates defined in \eqref{uvcompnonstand}. The extension allowed us to plot regions of spacetime in Figure \ref{diag_nonhomo_ones} that were not covered by the pair $(\lambda, x_0)$.

We point out that one can perform the analogous extension for the coordinates defined in \eqref{uvcomptrad}. Again, with that extension we were able to plot the singularity and the future null infinities in Figure \ref{diag_OS_ones}. However, while extending these coordinates \eqref{uvcomptrad} one has to remember about correct blocks' constants and the sings choices.

\section{Additional plots of conformal diagrams} \label{app_add_plots}

In order to allow for more comfort in interpreting quite cluttered diagrams presented in Figures~\ref{diag_OS} and~\ref{diag_nonhomo_ones} here we provide some supplementary plots. Since these figures contain numerous overlapping objects, these supplementary plots contain only fewer number of selected elements.
\onecolumngrid

\begin{figure*}[ht]
    \centering
    \includegraphics[width=1\linewidth]{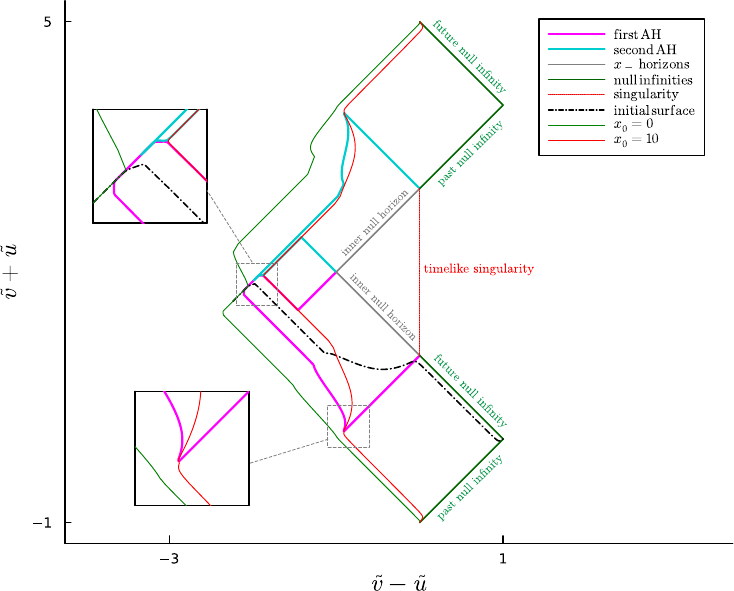}
    \caption{A version of the conformal diagram shown in Figure~\ref{diag_OS}, excluding the timelike geodesics labeled by $ x_0 \in \{5, 11, 15\} $. The geodesic labeled by \( x_0 = 0 \) follows the origin of the radial coordinate, i.e., \( x = 0 \) and the other geodesic with \( x_0 = 10 \) follows the surface of the homogeneous collapsing dust ball.
  }
    \label{dia_OS_notraj}
\end{figure*}

\clearpage

\begin{figure*}[ht]
    \centering
    \includegraphics[width=1\linewidth]{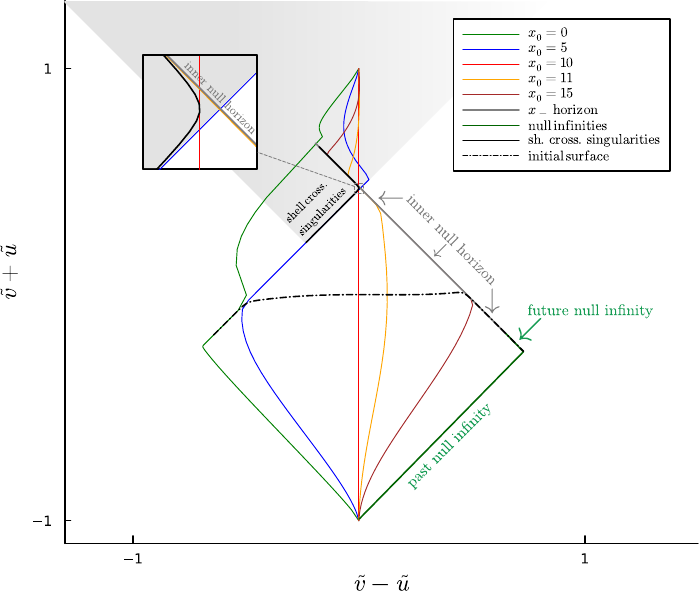}
    \caption{A version of the conformal diagram shown in Figure~\ref{diag_nonhomo_ones}, with the 'first AH' (the apparent horizon penetrable by timelike geodesics) omitted.
 }
    \label{dia_nonhomo_noappa}
\end{figure*}

\clearpage
\twocolumngrid

\bibliography{sample}

\begin{thebibliography}{35}%
\makeatletter
\providecommand \@ifxundefined [1]{%
 \@ifx{#1\undefined}
}%
\providecommand \@ifnum [1]{%
 \ifnum #1\expandafter \@firstoftwo
 \else \expandafter \@secondoftwo
 \fi
}%
\providecommand \@ifx [1]{%
 \ifx #1\expandafter \@firstoftwo
 \else \expandafter \@secondoftwo
 \fi
}%
\providecommand \natexlab [1]{#1}%
\providecommand \enquote  [1]{``#1''}%
\providecommand \bibnamefont  [1]{#1}%
\providecommand \bibfnamefont [1]{#1}%
\providecommand \citenamefont [1]{#1}%
\providecommand \href@noop [0]{\@secondoftwo}%
\providecommand \href [0]{\begingroup \@sanitize@url \@href}%
\providecommand \@href[1]{\@@startlink{#1}\@@href}%
\providecommand \@@href[1]{\endgroup#1\@@endlink}%
\providecommand \@sanitize@url [0]{\catcode `\\12\catcode `\$12\catcode `\&12\catcode `\#12\catcode `\^12\catcode `\_12\catcode `\%12\relax}%
\providecommand \@@startlink[1]{}%
\providecommand \@@endlink[0]{}%
\providecommand \url  [0]{\begingroup\@sanitize@url \@url }%
\providecommand \@url [1]{\endgroup\@href {#1}{\urlprefix }}%
\providecommand \urlprefix  [0]{URL }%
\providecommand \Eprint [0]{\href }%
\providecommand \doibase [0]{http://dx.doi.org/}%
\providecommand \selectlanguage [0]{\@gobble}%
\providecommand \bibinfo  [0]{\@secondoftwo}%
\providecommand \bibfield  [0]{\@secondoftwo}%
\providecommand \translation [1]{[#1]}%
\providecommand \BibitemOpen [0]{}%
\providecommand \bibitemStop [0]{}%
\providecommand \bibitemNoStop [0]{.\EOS\space}%
\providecommand \EOS [0]{\spacefactor3000\relax}%
\providecommand \BibitemShut  [1]{\csname bibitem#1\endcsname}%
\let\auto@bib@innerbib\@empty
\bibitem [{\citenamefont {Coogan}\ \emph {et~al.}(2021)\citenamefont {Coogan}, \citenamefont {Morrison},\ and\ \citenamefont {Profumo}}]{Coogan:2020tuf}%
  \BibitemOpen
  \bibfield  {author} {\bibinfo {author} {\bibfnamefont {Adam}\ \bibnamefont {Coogan}}, \bibinfo {author} {\bibfnamefont {Logan}\ \bibnamefont {Morrison}}, \ and\ \bibinfo {author} {\bibfnamefont {Stefano}\ \bibnamefont {Profumo}},\ }\bibfield  {title} {\enquote {\bibinfo {title} {{Direct Detection of Hawking Radiation from Asteroid-Mass Primordial Black Holes}},}\ }\href {\doibase 10.1103/PhysRevLett.126.171101} {\bibfield  {journal} {\bibinfo  {journal} {Phys. Rev. Lett.}\ }\textbf {\bibinfo {volume} {126}},\ \bibinfo {pages} {171101} (\bibinfo {year} {2021})},\ \Eprint {http://arxiv.org/abs/2010.04797} {arXiv:2010.04797 [astro-ph.CO]} \BibitemShut {NoStop}%
\bibitem [{\citenamefont {Hawking}(1975)}]{Hawking:1975vcx}%
  \BibitemOpen
  \bibfield  {author} {\bibinfo {author} {\bibfnamefont {S.~W.}\ \bibnamefont {Hawking}},\ }\bibfield  {title} {\enquote {\bibinfo {title} {{Particle Creation by Black Holes}},}\ }\href {\doibase 10.1007/BF02345020} {\bibfield  {journal} {\bibinfo  {journal} {Commun. Math. Phys.}\ }\textbf {\bibinfo {volume} {43}},\ \bibinfo {pages} {199--220} (\bibinfo {year} {1975})},\ \bibinfo {note} {[Erratum: Commun.Math.Phys. 46, 206 (1976)]}\BibitemShut {NoStop}%
\bibitem [{\citenamefont {Hawking}(1976)}]{Hawking:1976ra}%
  \BibitemOpen
  \bibfield  {author} {\bibinfo {author} {\bibfnamefont {S.~W.}\ \bibnamefont {Hawking}},\ }\bibfield  {title} {\enquote {\bibinfo {title} {{Breakdown of Predictability in Gravitational Collapse}},}\ }\href {\doibase 10.1103/PhysRevD.14.2460} {\bibfield  {journal} {\bibinfo  {journal} {Phys. Rev. D}\ }\textbf {\bibinfo {volume} {14}},\ \bibinfo {pages} {2460--2473} (\bibinfo {year} {1976})}\BibitemShut {NoStop}%
\bibitem [{\citenamefont {Mathur}(2009)}]{Mathur:2009hf}%
  \BibitemOpen
  \bibfield  {author} {\bibinfo {author} {\bibfnamefont {Samir~D.}\ \bibnamefont {Mathur}},\ }\bibfield  {title} {\enquote {\bibinfo {title} {{The Information paradox: A Pedagogical introduction}},}\ }\href {\doibase 10.1088/0264-9381/26/22/224001} {\bibfield  {journal} {\bibinfo  {journal} {Class. Quant. Grav.}\ }\textbf {\bibinfo {volume} {26}},\ \bibinfo {pages} {224001} (\bibinfo {year} {2009})},\ \Eprint {http://arxiv.org/abs/0909.1038} {arXiv:0909.1038 [hep-th]} \BibitemShut {NoStop}%
\bibitem [{\citenamefont {Ashtekar}\ and\ \citenamefont {Bojowald}(2005)}]{Ashtekar:2005cj}%
  \BibitemOpen
  \bibfield  {author} {\bibinfo {author} {\bibfnamefont {Abhay}\ \bibnamefont {Ashtekar}}\ and\ \bibinfo {author} {\bibfnamefont {Martin}\ \bibnamefont {Bojowald}},\ }\bibfield  {title} {\enquote {\bibinfo {title} {{Black hole evaporation: A Paradigm}},}\ }\href {\doibase 10.1088/0264-9381/22/16/014} {\bibfield  {journal} {\bibinfo  {journal} {Class. Quant. Grav.}\ }\textbf {\bibinfo {volume} {22}},\ \bibinfo {pages} {3349--3362} (\bibinfo {year} {2005})},\ \Eprint {http://arxiv.org/abs/gr-qc/0504029} {arXiv:gr-qc/0504029} \BibitemShut {NoStop}%
\bibitem [{\citenamefont {Bobula}()}]{BobulaLOOPS}%
  \BibitemOpen
  \bibfield  {author} {\bibinfo {author} {\bibfnamefont {Micha{\l}}\ \bibnamefont {Bobula}},\ }\href@noop {} {\enquote {\bibinfo {title} {{Radiation in quantum gravitational collapse}},}\ }\bibinfo {howpublished} {\url{ https://indico.cern.ch/event/1100970/}},\ \bibinfo {note} {{LOOPS’22 (18-22 July 2022), ENS de Lyon}}\BibitemShut {NoStop}%
\bibitem [{\citenamefont {Bobula}\ and\ \citenamefont {Paw\l{}owski}(2023)}]{Bobula:2023kbo}%
  \BibitemOpen
  \bibfield  {author} {\bibinfo {author} {\bibfnamefont {Micha\l{}}\ \bibnamefont {Bobula}}\ and\ \bibinfo {author} {\bibfnamefont {Tomasz}\ \bibnamefont {Paw\l{}owski}},\ }\bibfield  {title} {\enquote {\bibinfo {title} {{Rainbow Oppenheimer-Snyder collapse and the entanglement entropy production}},}\ }\href {\doibase 10.1103/PhysRevD.108.026016} {\bibfield  {journal} {\bibinfo  {journal} {Phys. Rev. D}\ }\textbf {\bibinfo {volume} {108}},\ \bibinfo {pages} {026016} (\bibinfo {year} {2023})},\ \Eprint {http://arxiv.org/abs/2303.12708} {arXiv:2303.12708 [gr-qc]} \BibitemShut {NoStop}%
\bibitem [{\citenamefont {Lewandowski}\ \emph {et~al.}(2023)\citenamefont {Lewandowski}, \citenamefont {Ma}, \citenamefont {Yang},\ and\ \citenamefont {Zhang}}]{Lewandowski:2022zce}%
  \BibitemOpen
  \bibfield  {author} {\bibinfo {author} {\bibfnamefont {Jerzy}\ \bibnamefont {Lewandowski}}, \bibinfo {author} {\bibfnamefont {Yongge}\ \bibnamefont {Ma}}, \bibinfo {author} {\bibfnamefont {Jinsong}\ \bibnamefont {Yang}}, \ and\ \bibinfo {author} {\bibfnamefont {Cong}\ \bibnamefont {Zhang}},\ }\bibfield  {title} {\enquote {\bibinfo {title} {{Quantum Oppenheimer-Snyder and Swiss Cheese Models}},}\ }\href {\doibase 10.1103/PhysRevLett.130.101501} {\bibfield  {journal} {\bibinfo  {journal} {Phys. Rev. Lett.}\ }\textbf {\bibinfo {volume} {130}},\ \bibinfo {pages} {101501} (\bibinfo {year} {2023})},\ \Eprint {http://arxiv.org/abs/2210.02253} {arXiv:2210.02253 [gr-qc]} \BibitemShut {NoStop}%
\bibitem [{\citenamefont {Han}\ \emph {et~al.}(2023)\citenamefont {Han}, \citenamefont {Rovelli},\ and\ \citenamefont {Soltani}}]{Han:2023wxg}%
  \BibitemOpen
  \bibfield  {author} {\bibinfo {author} {\bibfnamefont {Muxin}\ \bibnamefont {Han}}, \bibinfo {author} {\bibfnamefont {Carlo}\ \bibnamefont {Rovelli}}, \ and\ \bibinfo {author} {\bibfnamefont {Farshid}\ \bibnamefont {Soltani}},\ }\bibfield  {title} {\enquote {\bibinfo {title} {{Geometry of the black-to-white hole transition within a single asymptotic region}},}\ }\href {\doibase 10.1103/PhysRevD.107.064011} {\bibfield  {journal} {\bibinfo  {journal} {Phys. Rev. D}\ }\textbf {\bibinfo {volume} {107}},\ \bibinfo {pages} {064011} (\bibinfo {year} {2023})},\ \Eprint {http://arxiv.org/abs/2302.03872} {arXiv:2302.03872 [gr-qc]} \BibitemShut {NoStop}%
\bibitem [{\citenamefont {Giesel}\ \emph {et~al.}(2023)\citenamefont {Giesel}, \citenamefont {Han}, \citenamefont {Li}, \citenamefont {Liu},\ and\ \citenamefont {Singh}}]{Giesel:2022rxi}%
  \BibitemOpen
  \bibfield  {author} {\bibinfo {author} {\bibfnamefont {Kristina}\ \bibnamefont {Giesel}}, \bibinfo {author} {\bibfnamefont {Muxin}\ \bibnamefont {Han}}, \bibinfo {author} {\bibfnamefont {Bao-Fei}\ \bibnamefont {Li}}, \bibinfo {author} {\bibfnamefont {Hongguang}\ \bibnamefont {Liu}}, \ and\ \bibinfo {author} {\bibfnamefont {Parampreet}\ \bibnamefont {Singh}},\ }\bibfield  {title} {\enquote {\bibinfo {title} {{Spherical symmetric gravitational collapse of a dust cloud: Polymerized dynamics in reduced phase space}},}\ }\href {\doibase 10.1103/PhysRevD.107.044047} {\bibfield  {journal} {\bibinfo  {journal} {Phys. Rev. D}\ }\textbf {\bibinfo {volume} {107}},\ \bibinfo {pages} {044047} (\bibinfo {year} {2023})},\ \Eprint {http://arxiv.org/abs/2212.01930} {arXiv:2212.01930 [gr-qc]} \BibitemShut {NoStop}%
\bibitem [{\citenamefont {Giesel}\ \emph {et~al.}(2024{\natexlab{a}})\citenamefont {Giesel}, \citenamefont {Liu}, \citenamefont {Rullit}, \citenamefont {Singh},\ and\ \citenamefont {Weigl}}]{Giesel:2023tsj}%
  \BibitemOpen
  \bibfield  {author} {\bibinfo {author} {\bibfnamefont {Kristina}\ \bibnamefont {Giesel}}, \bibinfo {author} {\bibfnamefont {Hongguang}\ \bibnamefont {Liu}}, \bibinfo {author} {\bibfnamefont {Eric}\ \bibnamefont {Rullit}}, \bibinfo {author} {\bibfnamefont {Parampreet}\ \bibnamefont {Singh}}, \ and\ \bibinfo {author} {\bibfnamefont {Stefan~Andreas}\ \bibnamefont {Weigl}},\ }\bibfield  {title} {\enquote {\bibinfo {title} {{Embedding generalized Lema\^\i{}tre-Tolman-Bondi models in polymerized spherically symmetric spacetimes}},}\ }\href {\doibase 10.1103/PhysRevD.110.104017} {\bibfield  {journal} {\bibinfo  {journal} {Phys. Rev. D}\ }\textbf {\bibinfo {volume} {110}},\ \bibinfo {pages} {104017} (\bibinfo {year} {2024}{\natexlab{a}})},\ \Eprint {http://arxiv.org/abs/2308.10949} {arXiv:2308.10949 [gr-qc]} \BibitemShut {NoStop}%
\bibitem [{\citenamefont {Giesel}\ \emph {et~al.}(2024{\natexlab{b}})\citenamefont {Giesel}, \citenamefont {Liu}, \citenamefont {Singh},\ and\ \citenamefont {Weigl}}]{Giesel:2023hys}%
  \BibitemOpen
  \bibfield  {author} {\bibinfo {author} {\bibfnamefont {Kristina}\ \bibnamefont {Giesel}}, \bibinfo {author} {\bibfnamefont {Hongguang}\ \bibnamefont {Liu}}, \bibinfo {author} {\bibfnamefont {Parampreet}\ \bibnamefont {Singh}}, \ and\ \bibinfo {author} {\bibfnamefont {Stefan~Andreas}\ \bibnamefont {Weigl}},\ }\bibfield  {title} {\enquote {\bibinfo {title} {{Generalized analysis of a dust collapse in effective loop quantum gravity: Fate of shocks and covariance}},}\ }\href {\doibase 10.1103/PhysRevD.110.104016} {\bibfield  {journal} {\bibinfo  {journal} {Phys. Rev. D}\ }\textbf {\bibinfo {volume} {110}},\ \bibinfo {pages} {104016} (\bibinfo {year} {2024}{\natexlab{b}})},\ \Eprint {http://arxiv.org/abs/2308.10953} {arXiv:2308.10953 [gr-qc]} \BibitemShut {NoStop}%
\bibitem [{\citenamefont {Giesel}\ \emph {et~al.}(2024{\natexlab{c}})\citenamefont {Giesel}, \citenamefont {Liu}, \citenamefont {Singh},\ and\ \citenamefont {Weigl}}]{Giesel:2024mps}%
  \BibitemOpen
  \bibfield  {author} {\bibinfo {author} {\bibfnamefont {Kristina}\ \bibnamefont {Giesel}}, \bibinfo {author} {\bibfnamefont {Hongguang}\ \bibnamefont {Liu}}, \bibinfo {author} {\bibfnamefont {Parampreet}\ \bibnamefont {Singh}}, \ and\ \bibinfo {author} {\bibfnamefont {Stefan~Andreas}\ \bibnamefont {Weigl}},\ }\bibfield  {title} {\enquote {\bibinfo {title} {{Regular black holes and their relationship to polymerized models and mimetic gravity}},}\ }\href@noop {} {\  (\bibinfo {year} {2024}{\natexlab{c}})},\ \Eprint {http://arxiv.org/abs/2405.03554} {arXiv:2405.03554 [gr-qc]} \BibitemShut {NoStop}%
\bibitem [{\citenamefont {Kelly}\ \emph {et~al.}(2021)\citenamefont {Kelly}, \citenamefont {Santacruz},\ and\ \citenamefont {Wilson-Ewing}}]{Kelly:2020lec}%
  \BibitemOpen
  \bibfield  {author} {\bibinfo {author} {\bibfnamefont {Jarod~George}\ \bibnamefont {Kelly}}, \bibinfo {author} {\bibfnamefont {Robert}\ \bibnamefont {Santacruz}}, \ and\ \bibinfo {author} {\bibfnamefont {Edward}\ \bibnamefont {Wilson-Ewing}},\ }\bibfield  {title} {\enquote {\bibinfo {title} {{Black hole collapse and bounce in effective loop quantum gravity}},}\ }\href {\doibase 10.1088/1361-6382/abd3e2} {\bibfield  {journal} {\bibinfo  {journal} {Class. Quant. Grav.}\ }\textbf {\bibinfo {volume} {38}},\ \bibinfo {pages} {04LT01} (\bibinfo {year} {2021})},\ \Eprint {http://arxiv.org/abs/2006.09325} {arXiv:2006.09325 [gr-qc]} \BibitemShut {NoStop}%
\bibitem [{\citenamefont {Kelly}\ \emph {et~al.}(2020)\citenamefont {Kelly}, \citenamefont {Santacruz},\ and\ \citenamefont {Wilson-Ewing}}]{Kelly:2020uwj}%
  \BibitemOpen
  \bibfield  {author} {\bibinfo {author} {\bibfnamefont {Jarod~George}\ \bibnamefont {Kelly}}, \bibinfo {author} {\bibfnamefont {Robert}\ \bibnamefont {Santacruz}}, \ and\ \bibinfo {author} {\bibfnamefont {Edward}\ \bibnamefont {Wilson-Ewing}},\ }\bibfield  {title} {\enquote {\bibinfo {title} {{Effective loop quantum gravity framework for vacuum spherically symmetric spacetimes}},}\ }\href {\doibase 10.1103/PhysRevD.102.106024} {\bibfield  {journal} {\bibinfo  {journal} {Phys. Rev. D}\ }\textbf {\bibinfo {volume} {102}},\ \bibinfo {pages} {106024} (\bibinfo {year} {2020})},\ \Eprint {http://arxiv.org/abs/2006.09302} {arXiv:2006.09302 [gr-qc]} \BibitemShut {NoStop}%
\bibitem [{\citenamefont {Husain}\ \emph {et~al.}(2022{\natexlab{a}})\citenamefont {Husain}, \citenamefont {Kelly}, \citenamefont {Santacruz},\ and\ \citenamefont {Wilson-Ewing}}]{Husain:2021ojz}%
  \BibitemOpen
  \bibfield  {author} {\bibinfo {author} {\bibfnamefont {Viqar}\ \bibnamefont {Husain}}, \bibinfo {author} {\bibfnamefont {Jarod~George}\ \bibnamefont {Kelly}}, \bibinfo {author} {\bibfnamefont {Robert}\ \bibnamefont {Santacruz}}, \ and\ \bibinfo {author} {\bibfnamefont {Edward}\ \bibnamefont {Wilson-Ewing}},\ }\bibfield  {title} {\enquote {\bibinfo {title} {{Quantum Gravity of Dust Collapse: Shock Waves from Black Holes}},}\ }\href {\doibase 10.1103/PhysRevLett.128.121301} {\bibfield  {journal} {\bibinfo  {journal} {Phys. Rev. Lett.}\ }\textbf {\bibinfo {volume} {128}},\ \bibinfo {pages} {121301} (\bibinfo {year} {2022}{\natexlab{a}})},\ \Eprint {http://arxiv.org/abs/2109.08667} {arXiv:2109.08667 [gr-qc]} \BibitemShut {NoStop}%
\bibitem [{\citenamefont {Husain}\ \emph {et~al.}(2022{\natexlab{b}})\citenamefont {Husain}, \citenamefont {Kelly}, \citenamefont {Santacruz},\ and\ \citenamefont {Wilson-Ewing}}]{Husain:2022gwp}%
  \BibitemOpen
  \bibfield  {author} {\bibinfo {author} {\bibfnamefont {Viqar}\ \bibnamefont {Husain}}, \bibinfo {author} {\bibfnamefont {Jarod~George}\ \bibnamefont {Kelly}}, \bibinfo {author} {\bibfnamefont {Robert}\ \bibnamefont {Santacruz}}, \ and\ \bibinfo {author} {\bibfnamefont {Edward}\ \bibnamefont {Wilson-Ewing}},\ }\bibfield  {title} {\enquote {\bibinfo {title} {{Fate of quantum black holes}},}\ }\href {\doibase 10.1103/PhysRevD.106.024014} {\bibfield  {journal} {\bibinfo  {journal} {Phys. Rev. D}\ }\textbf {\bibinfo {volume} {106}},\ \bibinfo {pages} {024014} (\bibinfo {year} {2022}{\natexlab{b}})},\ \Eprint {http://arxiv.org/abs/2203.04238} {arXiv:2203.04238 [gr-qc]} \BibitemShut {NoStop}%
\bibitem [{\citenamefont {Hergott}\ \emph {et~al.}(2022)\citenamefont {Hergott}, \citenamefont {Husain},\ and\ \citenamefont {Rastgoo}}]{Hergott:2022hjm}%
  \BibitemOpen
  \bibfield  {author} {\bibinfo {author} {\bibfnamefont {Samantha}\ \bibnamefont {Hergott}}, \bibinfo {author} {\bibfnamefont {Viqar}\ \bibnamefont {Husain}}, \ and\ \bibinfo {author} {\bibfnamefont {Saeed}\ \bibnamefont {Rastgoo}},\ }\bibfield  {title} {\enquote {\bibinfo {title} {{Model metrics for quantum black hole evolution: Gravitational collapse, singularity resolution, and transient horizons}},}\ }\href {\doibase 10.1103/PhysRevD.106.046012} {\bibfield  {journal} {\bibinfo  {journal} {Phys. Rev. D}\ }\textbf {\bibinfo {volume} {106}},\ \bibinfo {pages} {046012} (\bibinfo {year} {2022})},\ \Eprint {http://arxiv.org/abs/2206.06425} {arXiv:2206.06425 [gr-qc]} \BibitemShut {NoStop}%
\bibitem [{\citenamefont {Fazzini}\ \emph {et~al.}(2023)\citenamefont {Fazzini}, \citenamefont {Rovelli},\ and\ \citenamefont {Soltani}}]{Fazzini:2023scu}%
  \BibitemOpen
  \bibfield  {author} {\bibinfo {author} {\bibfnamefont {Francesco}\ \bibnamefont {Fazzini}}, \bibinfo {author} {\bibfnamefont {Carlo}\ \bibnamefont {Rovelli}}, \ and\ \bibinfo {author} {\bibfnamefont {Farshid}\ \bibnamefont {Soltani}},\ }\bibfield  {title} {\enquote {\bibinfo {title} {{Painlev\'e-Gullstrand coordinates discontinuity in the quantum Oppenheimer-Snyder model}},}\ }\href {\doibase 10.1103/PhysRevD.108.044009} {\bibfield  {journal} {\bibinfo  {journal} {Phys. Rev. D}\ }\textbf {\bibinfo {volume} {108}},\ \bibinfo {pages} {044009} (\bibinfo {year} {2023})},\ \Eprint {http://arxiv.org/abs/2307.07797} {arXiv:2307.07797 [gr-qc]} \BibitemShut {NoStop}%
\bibitem [{\citenamefont {Fazzini}\ \emph {et~al.}(2024)\citenamefont {Fazzini}, \citenamefont {Husain},\ and\ \citenamefont {Wilson-Ewing}}]{Fazzini:2023ova}%
  \BibitemOpen
  \bibfield  {author} {\bibinfo {author} {\bibfnamefont {Francesco}\ \bibnamefont {Fazzini}}, \bibinfo {author} {\bibfnamefont {Viqar}\ \bibnamefont {Husain}}, \ and\ \bibinfo {author} {\bibfnamefont {Edward}\ \bibnamefont {Wilson-Ewing}},\ }\bibfield  {title} {\enquote {\bibinfo {title} {{Shell-crossings and shock formation during gravitational collapse in effective loop quantum gravity}},}\ }\href {\doibase 10.1103/PhysRevD.109.084052} {\bibfield  {journal} {\bibinfo  {journal} {Phys. Rev. D}\ }\textbf {\bibinfo {volume} {109}},\ \bibinfo {pages} {084052} (\bibinfo {year} {2024})},\ \Eprint {http://arxiv.org/abs/2312.02032} {arXiv:2312.02032 [gr-qc]} \BibitemShut {NoStop}%
\bibitem [{\citenamefont {Cipriani}\ \emph {et~al.}(2024)\citenamefont {Cipriani}, \citenamefont {Fazzini},\ and\ \citenamefont {Wilson-Ewing}}]{Cipriani:2024nhx}%
  \BibitemOpen
  \bibfield  {author} {\bibinfo {author} {\bibfnamefont {Lorenzo}\ \bibnamefont {Cipriani}}, \bibinfo {author} {\bibfnamefont {Francesco}\ \bibnamefont {Fazzini}}, \ and\ \bibinfo {author} {\bibfnamefont {Edward}\ \bibnamefont {Wilson-Ewing}},\ }\bibfield  {title} {\enquote {\bibinfo {title} {{Gravitational collapse in effective loop quantum gravity: Beyond marginally bound configurations}},}\ }\href {\doibase 10.1103/PhysRevD.110.066004} {\bibfield  {journal} {\bibinfo  {journal} {Phys. Rev. D}\ }\textbf {\bibinfo {volume} {110}},\ \bibinfo {pages} {066004} (\bibinfo {year} {2024})},\ \Eprint {http://arxiv.org/abs/2404.04192} {arXiv:2404.04192 [gr-qc]} \BibitemShut {NoStop}%
\bibitem [{\citenamefont {Wilson-Ewing}(2025)}]{Wilson-Ewing:2024uad}%
  \BibitemOpen
  \bibfield  {author} {\bibinfo {author} {\bibfnamefont {Edward}\ \bibnamefont {Wilson-Ewing}},\ }\bibfield  {title} {\enquote {\bibinfo {title} {{Static Planck stars from effective loop quantum gravity}},}\ }\href {\doibase 10.1209/0295-5075/adac08} {\bibfield  {journal} {\bibinfo  {journal} {EPL}\ }\textbf {\bibinfo {volume} {149}},\ \bibinfo {pages} {39002} (\bibinfo {year} {2025})},\ \Eprint {http://arxiv.org/abs/2408.16533} {arXiv:2408.16533 [gr-qc]} \BibitemShut {NoStop}%
\bibitem [{\citenamefont {Wilson-Ewing}(2024)}]{Wilson-Ewing:2024fxo}%
  \BibitemOpen
  \bibfield  {author} {\bibinfo {author} {\bibfnamefont {Edward}\ \bibnamefont {Wilson-Ewing}},\ }\bibfield  {title} {\enquote {\bibinfo {title} {{Dynamical homogenization in effective loop quantum cosmology}},}\ }\href@noop {} {\  (\bibinfo {year} {2024})},\ \Eprint {http://arxiv.org/abs/2409.18889} {arXiv:2409.18889 [gr-qc]} \BibitemShut {NoStop}%
\bibitem [{\citenamefont {M\"unch}(2021)}]{Munch:2021oqn}%
  \BibitemOpen
  \bibfield  {author} {\bibinfo {author} {\bibfnamefont {Johannes}\ \bibnamefont {M\"unch}},\ }\bibfield  {title} {\enquote {\bibinfo {title} {{Causal structure of a recent loop quantum gravity black hole collapse model}},}\ }\href {\doibase 10.1103/PhysRevD.104.046019} {\bibfield  {journal} {\bibinfo  {journal} {Phys. Rev. D}\ }\textbf {\bibinfo {volume} {104}},\ \bibinfo {pages} {046019} (\bibinfo {year} {2021})},\ \Eprint {http://arxiv.org/abs/2103.17112} {arXiv:2103.17112 [gr-qc]} \BibitemShut {NoStop}%
\bibitem [{\citenamefont {Alonso-Bardaji}\ and\ \citenamefont {Brizuela}(2024)}]{Alonso-Bardaji:2023qgu}%
  \BibitemOpen
  \bibfield  {author} {\bibinfo {author} {\bibfnamefont {Asier}\ \bibnamefont {Alonso-Bardaji}}\ and\ \bibinfo {author} {\bibfnamefont {David}\ \bibnamefont {Brizuela}},\ }\bibfield  {title} {\enquote {\bibinfo {title} {{Nonsingular collapse of a spherical dust cloud}},}\ }\href {\doibase 10.1103/PhysRevD.109.064023} {\bibfield  {journal} {\bibinfo  {journal} {Phys. Rev. D}\ }\textbf {\bibinfo {volume} {109}},\ \bibinfo {pages} {064023} (\bibinfo {year} {2024})},\ \Eprint {http://arxiv.org/abs/2312.15505} {arXiv:2312.15505 [gr-qc]} \BibitemShut {NoStop}%
\bibitem [{\citenamefont {Rovelli}\ and\ \citenamefont {Vidotto}(2024)}]{Rovelli:2024sjl}%
  \BibitemOpen
  \bibfield  {author} {\bibinfo {author} {\bibfnamefont {Carlo}\ \bibnamefont {Rovelli}}\ and\ \bibinfo {author} {\bibfnamefont {Francesca}\ \bibnamefont {Vidotto}},\ }\bibfield  {title} {\enquote {\bibinfo {title} {{Planck stars, White Holes, Remnants and Planck-mass quasi-particles. The quantum gravity phase in black holes' evolution and its manifestations}},}\ }\href@noop {} {\  (\bibinfo {year} {2024})},\ \Eprint {http://arxiv.org/abs/2407.09584} {arXiv:2407.09584 [gr-qc]} \BibitemShut {NoStop}%
\bibitem [{\citenamefont {Ashtekar}\ \emph {et~al.}(2006)\citenamefont {Ashtekar}, \citenamefont {Pawlowski},\ and\ \citenamefont {Singh}}]{Ashtekar:2006wn}%
  \BibitemOpen
  \bibfield  {author} {\bibinfo {author} {\bibfnamefont {Abhay}\ \bibnamefont {Ashtekar}}, \bibinfo {author} {\bibfnamefont {Tomasz}\ \bibnamefont {Pawlowski}}, \ and\ \bibinfo {author} {\bibfnamefont {Parampreet}\ \bibnamefont {Singh}},\ }\bibfield  {title} {\enquote {\bibinfo {title} {{Quantum Nature of the Big Bang: Improved dynamics}},}\ }\href {\doibase 10.1103/PhysRevD.74.084003} {\bibfield  {journal} {\bibinfo  {journal} {Phys. Rev. D}\ }\textbf {\bibinfo {volume} {74}},\ \bibinfo {pages} {084003} (\bibinfo {year} {2006})},\ \Eprint {http://arxiv.org/abs/gr-qc/0607039} {arXiv:gr-qc/0607039} \BibitemShut {NoStop}%
\bibitem [{\citenamefont {Domagala}\ and\ \citenamefont {Lewandowski}(2004)}]{Domagala:2004jt}%
  \BibitemOpen
  \bibfield  {author} {\bibinfo {author} {\bibfnamefont {Marcin}\ \bibnamefont {Domagala}}\ and\ \bibinfo {author} {\bibfnamefont {Jerzy}\ \bibnamefont {Lewandowski}},\ }\bibfield  {title} {\enquote {\bibinfo {title} {{Black hole entropy from quantum geometry}},}\ }\href {\doibase 10.1088/0264-9381/21/22/014} {\bibfield  {journal} {\bibinfo  {journal} {Class. Quant. Grav.}\ }\textbf {\bibinfo {volume} {21}},\ \bibinfo {pages} {5233--5244} (\bibinfo {year} {2004})},\ \Eprint {http://arxiv.org/abs/gr-qc/0407051} {arXiv:gr-qc/0407051} \BibitemShut {NoStop}%
\bibitem [{\citenamefont {Meissner}(2004)}]{Meissner:2004ju}%
  \BibitemOpen
  \bibfield  {author} {\bibinfo {author} {\bibfnamefont {Krzysztof~A.}\ \bibnamefont {Meissner}},\ }\bibfield  {title} {\enquote {\bibinfo {title} {{Black hole entropy in loop quantum gravity}},}\ }\href {\doibase 10.1088/0264-9381/21/22/015} {\bibfield  {journal} {\bibinfo  {journal} {Class. Quant. Grav.}\ }\textbf {\bibinfo {volume} {21}},\ \bibinfo {pages} {5245--5252} (\bibinfo {year} {2004})},\ \Eprint {http://arxiv.org/abs/gr-qc/0407052} {arXiv:gr-qc/0407052} \BibitemShut {NoStop}%
\bibitem [{\citenamefont {Giesel}\ \emph {et~al.}(2022)\citenamefont {Giesel}, \citenamefont {Li}, \citenamefont {Singh},\ and\ \citenamefont {Weigl}}]{Giesel:2021rky}%
  \BibitemOpen
  \bibfield  {author} {\bibinfo {author} {\bibfnamefont {Kristina}\ \bibnamefont {Giesel}}, \bibinfo {author} {\bibfnamefont {Bao-Fei}\ \bibnamefont {Li}}, \bibinfo {author} {\bibfnamefont {Parampreet}\ \bibnamefont {Singh}}, \ and\ \bibinfo {author} {\bibfnamefont {Stefan~Andreas}\ \bibnamefont {Weigl}},\ }\bibfield  {title} {\enquote {\bibinfo {title} {{Consistent gauge-fixing conditions in polymerized gravitational systems}},}\ }\href {\doibase 10.1103/PhysRevD.105.066023} {\bibfield  {journal} {\bibinfo  {journal} {Phys. Rev. D}\ }\textbf {\bibinfo {volume} {105}},\ \bibinfo {pages} {066023} (\bibinfo {year} {2022})},\ \Eprint {http://arxiv.org/abs/2112.13860} {arXiv:2112.13860 [gr-qc]} \BibitemShut {NoStop}%
\bibitem [{\citenamefont {Lee}(2013)}]{Lee2013}%
  \BibitemOpen
  \bibfield  {author} {\bibinfo {author} {\bibfnamefont {John~M.}\ \bibnamefont {Lee}},\ }\href {\doibase 10.1007/978-1-4419-9982-5} {\emph {\bibinfo {title} {Introduction to Smooth Manifolds}}},\ \bibinfo {edition} {2nd}\ ed.,\ \bibinfo {series} {Graduate Texts in Mathematics}, Vol.\ \bibinfo {volume} {218}\ (\bibinfo  {publisher} {Springer},\ \bibinfo {address} {New York},\ \bibinfo {year} {2013})\BibitemShut {NoStop}%
\bibitem [{\citenamefont {Schindler}\ and\ \citenamefont {Aguirre}(2018)}]{Schindler:2018wbx}%
  \BibitemOpen
  \bibfield  {author} {\bibinfo {author} {\bibfnamefont {J.~C.}\ \bibnamefont {Schindler}}\ and\ \bibinfo {author} {\bibfnamefont {A.}~\bibnamefont {Aguirre}},\ }\bibfield  {title} {\enquote {\bibinfo {title} {{Algorithms for the explicit computation of Penrose diagrams}},}\ }\href {\doibase 10.1088/1361-6382/aabce2} {\bibfield  {journal} {\bibinfo  {journal} {Class. Quant. Grav.}\ }\textbf {\bibinfo {volume} {35}},\ \bibinfo {pages} {105019} (\bibinfo {year} {2018})},\ \Eprint {http://arxiv.org/abs/1802.02263} {arXiv:1802.02263 [gr-qc]} \BibitemShut {NoStop}%
\bibitem [{\citenamefont {Poisson}(2009)}]{Poisson:2009pwt}%
  \BibitemOpen
  \bibfield  {author} {\bibinfo {author} {\bibfnamefont {Eric}\ \bibnamefont {Poisson}},\ }\href {\doibase 10.1017/CBO9780511606601} {\emph {\bibinfo {title} {{A Relativist's Toolkit: The Mathematics of Black-Hole Mechanics}}}}\ (\bibinfo  {publisher} {Cambridge University Press},\ \bibinfo {year} {2009})\BibitemShut {NoStop}%
\bibitem [{\citenamefont {Szekeres}\ and\ \citenamefont {Lun}(1999)}]{Szekeres:1995gy}%
  \BibitemOpen
  \bibfield  {author} {\bibinfo {author} {\bibfnamefont {Peter}\ \bibnamefont {Szekeres}}\ and\ \bibinfo {author} {\bibfnamefont {Anthony}\ \bibnamefont {Lun}},\ }\bibfield  {title} {\enquote {\bibinfo {title} {{What is a shell crossing singularity?}}}\ }\href {\doibase 10.1017/S0334270000011140} {\bibfield  {journal} {\bibinfo  {journal} {J. Austral. Math. Soc. B}\ }\textbf {\bibinfo {volume} {41}},\ \bibinfo {pages} {167--179} (\bibinfo {year} {1999})}\BibitemShut {NoStop}%
\bibitem [{\citenamefont {Bonanno}\ \emph {et~al.}(2023)\citenamefont {Bonanno}, \citenamefont {Khosravi},\ and\ \citenamefont {Saueressig}}]{Bonanno:2022jjp}%
  \BibitemOpen
  \bibfield  {author} {\bibinfo {author} {\bibfnamefont {Alfio}\ \bibnamefont {Bonanno}}, \bibinfo {author} {\bibfnamefont {Amir-Pouyan}\ \bibnamefont {Khosravi}}, \ and\ \bibinfo {author} {\bibfnamefont {Frank}\ \bibnamefont {Saueressig}},\ }\bibfield  {title} {\enquote {\bibinfo {title} {{Regular evaporating black holes with stable cores}},}\ }\href {\doibase 10.1103/PhysRevD.107.024005} {\bibfield  {journal} {\bibinfo  {journal} {Phys. Rev. D}\ }\textbf {\bibinfo {volume} {107}},\ \bibinfo {pages} {024005} (\bibinfo {year} {2023})},\ \Eprint {http://arxiv.org/abs/2209.10612} {arXiv:2209.10612 [gr-qc]} \BibitemShut {NoStop}%
\end{thebibliography}%

\end{document}